\documentstyle[12pt,preprint]{aastex}
\begin{document}

\title{Galaxy morphology in the rich cluster Abell 2390}

\author{J.B.Hutchings\altaffilmark{1}, A.Saintonge, D.Schade\altaffilmark{1},
D.Frenette}
\affil{Herzberg Insitute of Astrophysics,
National Research Council of Canada,\\ Victoria, B.C., Canada}
\altaffiltext{1}{Guest Observer, Canada France Hawaii Telescope,
which is operated by NRC of Canada, CNRS of France, and the University 
of Hawaii}

\begin{abstract}
We have analysed images of the field of A2390 obtained with the CFHT and
HST. The analysis fits models to bulge and disk components to several
hundred galaxies, with about equal samples from the cluster and field. 
We also have assessed and graded asymmetries in the images. The cluster 
galaxies are compared in different cluster locations and also compared 
with field galaxies. We find that the central old population galaxies are 
bulge-dominated, while disk systems have young populations and are found
predominantly in the outer cluster. S0 and bulgy disk galaxies are found
throughout, but concentrate in regions of substructure. Disks of cluster
blue galaxies are generally brighter and smaller than those in the field. 
We find that the cluster members have a higher proportion of interacting
galaxies than the field sample. Interactions in the cluster and in the 
field, as well as cluster infall, appear to inhibit star-formation in 
galaxies.

\end{abstract}


\section{Introduction}

   The well-known rich cluster Abell 2390 at z=0.23 was studied as part 
of the CNOC1 cosmology program (Carlberg et al 1996). A detailed discussion
of its spectroscopic properties from the CNOC1 database was published by
Abraham et al (1996). From spectra of some 250 cluster galaxies, Abraham
et al showed that infall into the cluster truncates star-formation, although
there are some `near-field' galaxies that have young populations, and may not
yet be affected by the cluster environment.

    The morphological structure of cluster galaxies is of interest in
further understanding the effects of cluster infall and assimilation
of galaxies. It is presumed that the high fraction of red central
galaxies with old stellar populations are formed by tidal events that
occur within the cluster, and giving them a bulge-dominated morphology. 
However, the CNOC1 imaging data were not used for a systematic
morphological study until now.

   In this paper, we have made use of a variety of additional imaging data
of the cluster. We have performed a morphological modelling
analysis of the cluster galaxies, and also of a field galaxy sample from the
same databases. In our discussion we discuss the connection between
morphology, cluster location, and spectra, and look for differences  
of properties in a similar sample of field galaxies.

   The data used are from four databases. 1) the original CFHT MOS 
images from the CNOC1 database. These cover all the galaxies which 
have spectra, so are complete. However, the image quality and sampling 
are not optimal. 2) CFHT OSIS images have better sampling and image 
quality, from better observing conditions and with a tip-tilt imaging 
system. However, these images were part of a narrow-band imaging 
program and were not deeply exposed in the continuum band.
3) Images with the CFHT AO system in visible wavelengths produced
still better sampled images with FWHM in the range 0.3 to 0.4".
4) Finally, we used HST WFPC2 images from our own and archival databases.
These have FWHM about 0.2-0.3". 

Table 1 summarizes the data and compares their coverage of the cluster.
While the MOS data have the worst resolution, they offer the most extensive
coverage of the cluster and have the highest signal level per pixel.
Thus, in what follows, we have used the MOS data for most of the discussion,
but have used the other datasets as checks on the morphology results.
The analysis does create a PSF that may vary across the image fields, and
uses them to calibrate the images, so that image resolution is not the
overriding consideration for the images. While the OSIS data have superior
resolution and sampling, their coverage of the cluster is less complete and
the signal levels are much lower, so that they are not as useful. The
PUEO and WFPC2 data are good quality checks on a limited subsample of galaxies.

\section{Measurements}

   The galaxy images were classified using the software described by
Schade et al (1996). Saintonge et al (2001) describe the analysis in
detail for new data on some CNOC clusters, which is exactly the process 
followed for this discussion, including definitions of all quantities. 
Thus, we note only that the program was run on all
data in Table 1. In the case of A2390, the model fits were all checked 
manually, rather than by an automatic checking process that was developed 
later for the larger datasets, based on the A2390 experience. In this
process, a uniform numerical criterion for model fit was applied in
determining which models fits to include in the discussion. 

While this eliminates poorly-defined models for faint or flawed images, it
raises the question of whether a significant population of galaxies
is rejected from our discussion. However, including lower quality model fits
appears to increase noise in our plots rather than introducing clear
systematic changes. More importantly, we applied the same criteria to
the cluster and field galaxies, so the comparison between the two groups
should not be affected.

   In deciding to use the MOS data only for our discussion, we did a
comparison of the results for the galaxies in common to the datasets. The
correspondence was good between all pairs of measurements for galaxies
whose data quality flags were good in both. The overall comparison of the
principal measure (bulge/total light, or B/T) shows a mean difference of
0.02 over the range 0 to 1, with $\sigma$ of 0.37, for all galaxies covered
by more than one database. We have used the MOS data results restricted to
only the good quality galaxy measurements. This amounts to 161 cluster
galaxies, (of a total of 209 with spectroscopic measures) spread over a 
wide range of cluster positions (see Figure 1).

  The process of fitting parametric models to these galaxies includes, as 
the first step, `symmetrizing' the image to be fit. The galaxy image is 
rotated by 180$^o$ and subtracted from itself. The residuals are then clipped 
to leave only those portions of the residuals that are positive and that
deviate by more than 2$\sigma$ from symmetry. These high-significance
residuals are then subtracted from the original image leaving a `symmetric'
image. This symmetric image is then fit by the models.

   This process is very effective at removing the effects of companions and
asymmetric features on the model-fitting process. It does not alter the
possibility of separately classifying galaxy images as symmetric or 
asymmetric, and does not make classifications of interaction  probability 
less effective. In fact, the asymmetric parts of the image were used in
deriving the interaction index, described below.

   A central question is whether many of the interacting galaxies have been
removed from the sample because we are presenting only `good' fits.  While
there may be a few such galaxies that have been rejected, the symmetrizing
process makes it less likely that we have rejected very many moderately (or
less) interacting galaxies. We stress again that the comparison between
field and cluster is done with the same criteria for inclusion.

  We combined two field-galaxy samples for comparison: galaxies from the A2390
images with redshifts (mostly in the range 0.16 to .35, so comparable with the 
cluster redshift of 0.23); and field galaxies in the redshift range 0.19
to 0.27 using MOS images from other CNOC clusters. These datasets contained 97
and 111 galaxies, respectively.
Almost all the cluster and field galaxies in our morphological database
have spectra from the CNOC1 program, and we have thus combined the
morphological and spectroscopic measurements from Abraham et al (1996) in 
some of the discussions that follow.

   We examined the best (MOS and OSIS) galaxy images (and their
asymmetries as isolated by rotation) and classified their `interaction
level', based on the distinctness of tidal tails, connecting luminosity,
warped disks, off-centre nuclei, and other asymmetries, with reference
to a grid of models of galaxy interactions computed with a tree code. 
This classification was done
independently by two of us for a subset of galaxies to standardize the 
criteria, and the results are given for the full set on a 5-point scale. 
A similar classification was described in more detail by Hutchings and
Neff (1991).

   Finally, we made a measure of the local galaxy density by counting the
distance to the 10th-nearest neighbour galaxies on the sky. The density
value used was proportional to the log of the inverse square of this radius. 
This enabled us to investigate the effects of local galaxy density 
independently of clustocentric radius (although without correction for
non-member galaxies).

Tables 2 and 3 show the values used in this paper for cluster and field
galaxies. In the sections below we present and discuss the more significant
correlations among the morphology, spectroscopic, photometric, and
positional measures of the cluster and field galaxies. Where sizes and
luminosities are shown, we have used H$_0$=50 and q$_0$=0.5.

\section{Cluster galaxy morphologies}

    Abraham et al (1996) discussed the spectroscopic and photometric
properties of the cluster in considerable detail. The diagrams in the present
paper illustrate the morphological parameters of the cluster galaxies,
as they relate to the spectroscopic and photometric data, and to their 
position in the cluster. The principal properties we discuss are the
bulge to disk flux ratio, and some aspects of their sizes and luminosities.

    Figure 1 shows the spatial distribution of the cluster galaxies in 
the sample. We have in general divided the galaxies into 3 types: `bulge',
which has modelled bulge/total light (B/T) of 0.8 or larger; `disk' which 
has B/T less than 0.4, and `S0', which has B/T between 0.4 and 0.8. The
intermediate classification probably contains all S0 type galaxies, but 
will also include others such as Sa with a bright bulge. The sample 
extends to large distances from the centre, along the principal axis of
extension of the cluster. The top panel shows the MOS results which we
have used, as discussed above. The lower panel shows the distribution
of the galaxies in the other datasets analysed. The OSIS data have gaps
at intermediate radii, and are much noisier, while the PUEO and HST
data cover only the central region.

   Figure 2 shows the spatial distribution of B/T with normalised radius
within the cluster. (The normalised radius is defined as the radius within
the cluster as a fraction of R200, the radius at which the cluster density 
is 200 times the critical density.) It also shows the same for the field 
sample, plotted with redshift mainly as a way to spread the points out 
(but does also show that we have no strong redshift-dependent effects). 
The field galaxies bracket the cluster redshift of 0.23 in any case.

The gap in the B/T distribution is caused by an explicit bias. In those
cases where a single component model (e.g. pure disk or pure bulge) is
indistinguishable statistically from a two-component model (bulge plus disk)
then we adopt the single component model. In the region B/T $>$ 0.8 (roughly)
the bulge clearly dominates the luminosity profile and it is frequently
difficult to detect a disk even if one is present (as determined from
simulations). Furthermore, the measured parameters of the bulge are affected
very little by the choice of pure bulge or bulge-plus-disk model. In these
cases too, we choose the single-component model.

The size of the gap in the B/T diagram depends on signal-to-noise ratio. If
the obserations all had very high S/N then there would be no gap because it
would be possible to detect even very weak disks in the presence of dominant
bulges.

 In Figure 2, there are clearly
separate populations of bulge galaxies that are not a continuation of 
the trends seen in the pure disk through increasing bulge up to B/T 0.8. 
This bulge population is more marked in the cluster, as expected, but galaxies
are found at all projected radii within the cluster. There are more
disk-dominated galaxies in the field, also as expected.

  In Figure 3 we show the distributions of intrinsic colour with disk surface
brightness for the two samples, which differ significantly. The
dotted lines mark the mean values from the field sample, which are fairly
evenly spread about these values. The greater spread to bluer colours
reflect the redshift range, which limits the observed red colour of the
galaxies. The cluster galaxy distribution is mainly skewed by the cluster red 
sequence, but also shows a paucity of low surface brightness disks
among the blue galaxies.

   In Figure 4 we compare the disk scale-lengths. These are related to
the absolute magnitude, with brighter galaxies having bigger disks. 
The dotted line is the linear fit to the log-log plot for field galaxies.
The cluster galaxies distribution differs from this as illustrated by the
dashed curve, which is a second order fit to the points: the disks are 
generally smaller (as well as brighter as seen in Figure 3), but there is 
a tail of larger disks that are the brightest ones in Figure 3.

    Figure 5 shows the distribution of galaxy types plotted with local
galaxy density. This is strongly correlated with radius within the cluster,
as shown by the radius values in the plot: the density is highest in the
cluster centre and falls with radius in the inner cluster. Beyond that there
is a region where radius and density appear largely unrelated. The outer
cluster shows a clear monotonic decline in density. Figure 5 has bins that
separate these three regions, with the central region divided into two 
bins of equal numbers of member galaxies. The plot shows what is generally
expected: the central regions strongly dominated by bulge galaxies, and a 
rising disk fraction with radius. The `S0' population remains near to 20\% at
all radii.

   Figure 6 compares some key spectroscopic measures with the B/T ratio.
These are the stellar population indicators used by Abraham et al (1996).
We have included all galaxies with good morphology measures but they
are not filtered by spectroscopic data quality, since these are not
readily available for the assembled field sample. The D4000 data are, however,
separated for the brightest galaxies, which will have better spectroscopic
data as well. There are some differences, as follows.

The [O II] emission, a signature of hot star formation, is somewhat
suppressed compared with the field galaxies, but has a wider range.
The differences are larger for lower B/T values.
H$\delta$ absorption, a post-starburst measure, is stronger for the cluster
galaxies, again more noticeably for the lower B/T values.
D4000, a measure of old stellar populations, is larger for the cluster 
galaxies, and more so for the intermediate B/T values. All these are consistent
with the idea proposed by Abraham et al (1996) for this cluster, and Morris
et al (1998) for another CNOC cluster, namely that infall into the cluster
truncates star-formation.

   Figure 7 plots the spectroscopic indices with our estimates of interaction
strength for cluster and field galaxies.  This plot shows the overall
differences discussed for Figure 6, and also illustrates connections to
disturbed galaxy morphology. The distributions in Figure 7 show significant 
difference beteeen cluster and field galaxies in all comparisions
The numbers on the right show the differences in percent of each sample
in each interactions subset: the cluster galaxies appear to be more
interacting than the field sample, again as expected. 

   In the spectroscopic measures in Figure 7, [O II] emission is stronger 
in the field sample, but both samples show decreased emission for stronger
interactions. This suggests that interactions inhibit star-formation
in all environments that are covered. H$\delta$ absorption is also stronger
in field galaxies, increasing with tidal interaction. The cluster galaxies
show a small trend the other way --- less H$\delta$ in highly interacting
systems.
The D4000 index decreases with interaction in both samples, but there is a
strong difference between field and cluster galaxies.

\section{Discussion}

    We find that the morphology measurements of A2390 cluster galaxies
generally support the expectations of the spectroscopic analysis by
Abraham et al (1996). The red central galaxies are bulge-dominated
systems, as expected from their old stellar populations. Such galaxies
are strongly concentrated in the cluster core. Disk systems on the other
hand are increasingly dominant in outer parts of the cluster. The S0
(and other bulgy disks) are found at all radii, but seem to be associated
with regions of subclumps within the cluster. This is consistent with
their being stripped disk galaxies as a result of recent infall into 
the cluster. 50\% of the field galaxies are disk-dominated (B/T=0),
while the cluster galaxies are dominated by the bulge-only (B/T=1)
population. The distribution of intermediate B/T values is similar for
the two samples, but they may refer to different morphologies (S0 and Sa, say)
which we cannot distinguish.

   The cluster disks are mostly fainter and smaller than the field galaxy
disks, which is presumably also a result of tidal or stripping effects
of infall. The few bright large disks are found in galaxies spread across
the cluster, with no particular other properties.

  The spectroscopic properties (Figures 6 and 7) indicate that the
disk-dominated systems have post-starburst stellar populations, compared
with the field sample, while the bulge-dominated galaxies are very similar
in cluster and field. This too supports the conclusion that infall into
the cluster truncates star-formation. The interaction statistics add some
more information to this: interactions are more common among the cluster
galaxies, as expected, but in both field and cluster samples, the [O II]
is increasingly suppressed with increased interaction visibility. Thus,
this truncation of star-formation appears to happen in non-cluster
environments too (although significantly more marked in the cluster).

  If we regard the interaction visibility as an indicator of the age of a
tidal event, the H$\delta$ data suggest a sequence of post-starburst
populations, again with the effects more marked in the cluster sample.

   Abraham et al (1996) noted a few near-field galaxies that appear to
be in the early stages of infall. These are blue and have young stellar
populations. Our morphology measures for these galaxies show them to be
disk-dominated, but with other properties that are varied but typical of 
the distribution of cluster galaxies. We have thus simply included them 
as cluster galaxies in our diagrams.

We are grateful to M. Balogh for making the OSIS images available to us
while still proprietary.

\clearpage

\begin{deluxetable}{lcccccrr}
\tablecaption{Comparison of the different data sets}
\tablehead{
\colhead{Data} & \colhead{Field of view} & \colhead{FWHM} &
\colhead{Pixel size} & \colhead{Exp time} &
\colhead{Band} & \colhead{\#galx\tablenotemark{1}}
&\colhead{sig/pixel\tablenotemark{2}}\\
&\colhead{(\# fields)} &\colhead{(")} &\colhead{(")} &\colhead{(sec)}  }

\startdata
MOS &$10'\times10'$ (5) &0.99 &0.315 &900 &R &236 &1160\\
OSIS &$4.1'\times3.4'$ (19) &0.64 &0.155 &90 &R &173 &30\\
PUEO &$2'\times2'$ (1) &0.43 &0.061 &1200x3 &I &13 &170\\
WFPC2 &$2.5'\times2.5'$ (3) &0.18 &0.105 &2100x5 &(I) &26 &350\\
\enddata
\tablenotetext{1}{Number of galaxies with morphology measurements}
\tablenotetext{2}{Proportional to the product of telescope throughput,
exposure, and pixel area on sky.}
\end{deluxetable}

\begin{deluxetable}{llllclrrrrrll}
\scriptsize
\tablecaption{Cluster galaxies}
\tablehead{
\colhead{z} & \colhead{$M_{B}$} & \colhead{$r_{norm}$} & \colhead{(U-V)$_{0}$} 
&\colhead{$B/T$} & \colhead{$\mu_{disk}$} & \colhead{$\mu_{bulge}$} 
&\colhead{Scl.ht}
&\colhead{1/2lt rad} &\colhead{$[O II]$} & \colhead{$H_{\delta}$} & \colhead{D4000} 
& \colhead{Int}\\
&\colhead{} &\colhead{(Mpc)} &&&&&\colhead{log(kpc)} &\colhead{log(kpc)}
&\colhead{(\AA)} &\colhead{(\AA)} &\colhead{} &\colhead{level}\\ }
\startdata
0.2316  &-18.70  &0.04  &1.88  &1.  &\nodata &12.2  &\nodata &0.30  &3.4  &-0.9  &2.16  &0.  \\
0.2238  &-19.45  &0.04  &1.98  &0.56  &19.8  &14.6  &0.29  &0.69  &4.8  &3.4  &2.50  &3.  \\
0.2150  &-19.66  &0.04  &2.04  &0.59  &19.7  &12.0  &0.30  &0.22  &-1.5  &0.8  &2.62  &2.  \\
0.2284  &-19.42  &0.04  &1.99  &1.  &\nodata &11.6  &\nodata &0.25  &0.6  &-1.0  &2.87  &\nodata  \\
0.2178  &-18.00  &0.04  &2.21  &0.  &20.9  &\nodata &0.36  &\nodata &-14.9  &2.8  &1.87  &\nodata  \\
0.2304  &-18.33  &0.04  &1.99  &1.  &\nodata &12.2  &\nodata &0.13  &3.2  &2.9  &2.49  &1.  \\
0.2237  &-17.92  &0.05  &2.01  &1.  &\nodata &12.9  &\nodata &0.35  &-5.1  &-6.4  &1.64  &\nodata  \\
0.2551  &-20.20  &0.05  &1.80  &0.69  &24.0  &13.3  &1.27  &0.69  &2.5  &0.7  &2.51  &\nodata  \\
0.2203  &-20.15  &0.06  &1.89  &0.56  &22.4  &12.5  &0.94  &0.41  &6.9  &1.0  &2.51  &0.  \\
0.2470  &-19.08  &0.07  &1.91  &0.27  &19.9  &12.8  &0.35  &0.09  &1.2  &-2.9  &2.20  &1.  \\
0.2188  &-19.48  &0.08  &1.95  &0.36  &20.8  &11.2  &0.58  &-0.07  &-6.0  &-2.8  &2.81  &0.  \\
0.2303  &-18.56  &0.08  &1.77  &1.  &\nodata &11.9  &\nodata &0.25  &1.4  &-2.0 &2.19  &1.  \\
0.2192  &-19.85  &0.08  &1.92  &1.  &\nodata &12.8  &\nodata &0.61  &3.6  &-1.2  &2.94  &3.  \\
0.2244  &-18.05  &0.09  &1.83  &0.52  &22.0  &9.7  &0.47  &-0.58  &5.2  &1.4  &2.38  &0.  \\
0.2292  &-18.63  &0.10  &1.77  &1.  &\nodata &12.1  &\nodata &0.49  &-3.0  &3.1  &2.08  &0.  \\
0.2214  &-19.26  &0.11  &1.93  &1.  &\nodata &12.7  &\nodata &0.46  &14.4  &0.2  &2.65  &\nodata  \\
0.2461  &-18.68  &0.11  &1.84  &1.  &\nodata &13.5  &\nodata &0.50  &0.1  &4.1  &2.52  &4.  \\
0.2287  &-19.27  &0.12  &1.84  &1.  &\nodata &13.3  &\nodata &0.67  &0.9  &2.8  &2.50  &\nodata  \\
0.2363  &-19.29  &0.12  &1.82  &0.22  &21.0  &8.1  &0.64  &-0.82  &-0.2  &0.8  &2.67  &\nodata  \\
0.2221  &-19.11  &0.13  &1.83  &1.  &\nodata &12.0  &\nodata &0.61  &6.5  &0.9  &2.53  &\nodata  \\
0.2288  &-19.61  &0.14  &2.04  &1.  &\nodata &13.2  &\nodata &0.79  &-0.1  &2.5  &2.07  &2.  \\
0.2199  &-18.79  &0.14  &1.24  &1.  &\nodata &11.1  &\nodata &0.29  &-1.0  &6.3  &1.92  &1.  \\
0.2422  &-18.48  &0.14  &1.89  &0.39  &20.8  &12.4  &0.37  &-0.02  &-3.9  &1.4  &2.29  &4.  \\
0.2281  &-19.83  &0.14  &1.92  &1.  &\nodata &11.9  &\nodata &0.67  &4.9  &-1.9  &2.67  &0.  \\
0.2294  &-19.24  &0.16  &1.90  &1.  &\nodata &12.7  &\nodata &0.75  &6.9  &-1.2  &2.28  &2.  \\
0.2313  &-19.97  &0.18  &1.90  &0.24  &20.6  &7.3  &0.68  &-0.84  &-4.6  &-1.2  &3.30  &0.  \\
0.2302  &-18.10  &0.18  &1.94  &0.51  &23.4  &14.0  &0.83  &0.34  &-26.8  &7.5  &2.31  &0.  \\
0.2238  &-19.32  &0.19  &2.01  &0.63  &21.2  &11.7  &0.51  &0.12  &3.6  &-1.2  &2.87  &0.  \\
0.2221  &-18.26  &0.20  &1.56  &1.  &\nodata &12.5  &\nodata &0.39  &-6.3  &5.6  &2.39  &\nodata  \\
0.2253  &-20.15  &0.20  &1.59  &0.2  &20.5  &7.2  &0.72  &-0.85  &1.1  &5.6  &2.01  &3.  \\
0.2319  &-19.01  &0.21  &1.92  &0.57  &20.5  &13.9  &0.34  &0.49  &-6.2  &1.2  &2.48  &\nodata  \\
0.2322  &-20.23  &0.22  &1.93  &0.44  &21.8  &12.2  &0.94  &0.33  &4.6  &-0.8  &2.63  &\nodata  \\
0.2328  &-18.23  &0.22  &1.80  &0.28  &20.4  &8.8  &0.31  &-0.84  &-14.7  &1.5  &1.88  &0.  \\
0.2324  &-18.88  &0.22  &1.86  &0.34  &20.7  &11.5  &0.45  &-0.14  &4.8  &-4.6  &2.43  &\nodata  \\
0.2354  &-18.97  &0.24  &1.85  &0.1  &21.6  &12.1  &0.71  &-0.28  &4.9  &0.4  &1.88  &0.  \\
0.2305  &-19.52  &0.24  &1.86  &1.  &\nodata &13.5  &\nodata &0.65  &0.5  &-2.5  &2.61  &\nodata  \\
0.2251  &-19.67  &0.24  &1.95  &1.  &\nodata &12.6  &\nodata &0.50  &-9.0  &-1.3  &3.26  &\nodata  \\
0.2294  &-18.85  &0.24  &1.91  &0.75  &21.4  &15.5  &0.41  &0.85  &-25.7  &-5.4  &2.86  &3.  \\
0.2290  &-18.60  &0.24  &1.16  &0.44  &20.8  &13.3  &0.41  &0.25  &-9.6  &8.3  &2.13  &1.  \\
0.2295  &-18.91  &0.25  &1.90  &0.46  &21.2  &9.8  &0.52  &-0.41  &11.3  &4.0  &2.33  &\nodata  \\
0.2224  &-17.79  &0.26  &1.85  &1.  &\nodata &14.1  &\nodata &0.45  &3.5  &3.1  &1.85  &3.  \\
0.2237  &-19.61  &0.27  &1.92  &0.29  &20.5  &8.8  &0.57  &-0.58  &16.9  &1.1  &2.18  &\nodata  \\
0.2245  &-19.77  &0.27  &1.78  &0.25  &20.8  &8.3  &0.69  &-0.66  &4.6  &-0.6  &2.67  &\nodata  \\
0.2318  &-18.62  &0.28  &1.95  &0.24  &20.9  &8.7  &0.46  &-0.84  &9.0  &3.2  &2.58  &3.  \\
0.2246  &-17.92  &0.29  &2.06  &1.  &\nodata &13.0  &\nodata &0.28  &8.4  &4.2  &3.09  &2.  \\
0.2318  &-18.89  &0.29  &1.85  &0.7  &23.4  &13.2  &0.83  &0.37  &13.1  &6.8  &2.14  &2.  \\
0.2202  &-18.95  &0.30  &1.92  &1.  &\nodata &11.7  &\nodata &0.45  &10.3  &2.1  &2.04  &3.  \\
0.2283  &-19.32  &0.30  &1.50  &0.  &20.6  &\nodata &0.64  &\nodata &11.0  &1.4  &2.49  &2.  \\
0.2251  &-18.15  &0.30  &1.33  &0.  &20.4  &\nodata &0.35  &\nodata &-4.4  &-6.9  &1.43  &1.  \\
0.2253  &-20.76  &0.30  &1.85  &1.  &\nodata &14.3  &\nodata &1.11  &-6.9  &-1.7  &2.72  &\nodata  \\
0.2237  &-18.99  &0.31  &1.95  &0.16  &21.1  &8.9  &0.58  &-0.83  &-5.8  &-0.5  &2.44  &0.  \\
0.2350  &-19.24  &0.32  &1.90  &1.  &\nodata &14.3  &\nodata &0.83  &20.1  &7.6  &2.44  &\nodata  \\
0.2285  &-19.19  &0.32  &1.71  &0.4  &19.9  &7.6  &0.31  &-0.84  &0.1  &2.6  &2.35  &\nodata  \\
0.2300  &-19.11  &0.32  &1.94  &0.36  &20.4  &8.2  &0.44  &-0.74  &14.1  &-0.5  &2.36  &1.  \\
0.2211  &-18.09  &0.32  &1.67  &0.  &22.8  &\nodata &0.83  &\nodata &-46.1  &12.2  &1.44  &1.  \\
0.2235  &-20.02  &0.32  &1.78  &0.27  &20.8  &7.2  &0.72  &-0.83  &-5.4  &0.9  &2.45  &0.  \\
0.2226  &-19.59  &0.33  &1.86  &1.  &\nodata &11.9  &\nodata &0.57  &-12.0  &2.3  &2.50  &\nodata  \\
0.2265  &-18.28  &0.33  &1.86  &1.  &\nodata &12.0  &\nodata &0.15  &1.2  &3.2  &2.21  &\nodata  \\
0.2275  &-18.55  &0.33  &1.75  &1.  &\nodata &12.8  &\nodata &0.68  &4.2  &-3.5  &3.43  &1.  \\
0.2292  &-18.43  &0.33  &1.93  &0.29  &21.3  &8.7  &0.47  &-0.84  &16.5  &4.7  &2.46  &0.  \\
0.2314  &-17.12  &0.34  &1.14  &0.42  &24.1  &12.7  &0.92  &-0.03  &58.7  &-0.7  &3.70  &0.  \\
0.2293  &-18.94  &0.34  &1.63  &0.  &20.3  &\nodata &0.52  &\nodata &-6.2  &-0.7  &1.89  &\nodata  \\
0.2248  &-18.59  &0.35  &1.52  &0.12  &20.3  &9.4  &0.40  &-0.84  &2.3  &3.9  &1.99  &\nodata  \\
0.2220  &-19.58  &0.35  &1.92  &0.3  &21.5  &10.1  &0.73  &-0.33  &-2.9  &3.7  &2.12  &\nodata  \\
0.2341  &-19.26  &0.35  &1.84  &1.  &\nodata &11.0  &\nodata &0.24  &4.6  &1.7  &2.25  &\nodata  \\
0.2433  &-17.70  &0.35  &1.76  &1.  &\nodata &11.8  &\nodata &0.09  &-4.5  &6.8  &2.13  &1.  \\
0.2313  &-17.87  &0.36  &1.87  &1.  &\nodata &13.2  &\nodata &0.43  &50.5  &6.9  &3.92  &0.  \\
0.2287  &-18.59  &0.36  &1.68  &1.  &\nodata &13.4  &\nodata &0.55  &-13.5  &-0.4  &2.59  &\nodata  \\
0.2253  &-19.19  &0.36  &1.87  &0.53  &21.8  &10.7  &0.64  &-0.15  &20.5  &-2.6  &3.34  &2.  \\
0.2166  &-19.32  &0.36  &1.99  &0.21  &20.4  &8.4  &0.47  &-0.84  &4.5  &-0.2  &2.53  &\nodata  \\
0.2388  &-18.63  &0.36  &1.45  &0.  &19.5  &\nodata &0.27  &\nodata &3.9  &4.3  &2.07  &0.  \\
0.2322  &-19.56  &0.36  &2.02  &0.51  &22.4  &11.8  &0.87  &0.14  &4.1  &-0.7  &2.65  &\nodata  \\
0.2269  &-19.71  &0.37  &1.93  &0.52  &22.9  &12.5  &0.99  &0.32  &12.0  &0.8  &3.05  &1.  \\
0.2158  &-19.38  &0.38  &1.74  &0.18  &21.1  &11.8  &0.67  &-0.12  &3.5  &2.2  &2.32  &3.  \\
0.2253  &-17.40  &0.38  &1.15  &0.  &20.2  &\nodata &0.16  &\nodata &-17.7  &4.8  &1.59  &3.  \\
0.2266  &-18.62  &0.39  &1.78  &1.  &\nodata &13.5  &\nodata &0.50  &12.6  &2.1  &2.09  &\nodata  \\
0.2311  &-18.76  &0.40  &1.90  &0.51  &21.7  &13.2  &0.57  &0.27  &-2.8  &-2.6  &2.40  &0.  \\
0.2409  &-18.41  &0.40  &1.89  &0.38  &21.6  &11.7  &0.53  &-0.17  &-27.1  &-3.8  &2.18  &0.  \\
0.2245  &-18.46  &0.40  &2.04  &0.39  &21.6  &12.9  &0.51  &0.07  &-3.1  &2.6  &3.22  &3.  \\
0.2321  &-18.44  &0.41  &1.90  &0.35  &20.5  &11.3  &0.33  &-0.25  &16.6  &0.6  &2.34  &\nodata  \\
0.2247  &-18.72  &0.43  &1.26  &0.  &19.8  &\nodata &0.37  &\nodata &9.3  &7.1  &1.84  &\nodata  \\
0.2427  &-18.16  &0.44  &0.85  &0.07  &21.3  &10.1  &0.59  &-0.84  &0.7  &3.8  &1.62  &2.  \\
0.2282  &-18.39  &0.47  &1.77  &1.  &\nodata &13.0  &\nodata &0.64  &7.4  &-0.8  &2.69  &4.  \\
0.2223  &-18.93  &0.51  &1.93  &0.3  &21.1  &11.5  &0.57  &-0.16  &37.5  &-2.0  &3.32  &\nodata  \\
0.2366  &-17.30  &0.51  &1.52  &0.  &21.2  &\nodata &0.36  &\nodata &7.8  &2.0  &2.14  &0.  \\
0.2280  &-18.19  &0.53  &1.81  &0.  &19.7  &\nodata &0.21  &\nodata &6.9  &4.8  &3.08  &2.  \\
0.2237  &-18.91  &0.58  &1.76  &1.  &\nodata &12.6  &\nodata &0.39  &0.9  &1.0  &2.40  &0.  \\
0.2313  &-18.89  &0.60  &1.82  &0.33  &21.8  &12.8  &0.71  &0.15  &-13.2  &3.7  &1.77  &\nodata  \\
0.2302  &-18.90  &0.60  &1.14  &0.  &20.2  &\nodata &0.49  &\nodata &-17.7  &2.5  &1.46  &1.  \\
0.2217  &-19.86  &0.62  &1.44  &0.22  &18.5  &8.3  &0.28  &-0.65  &-3.7  &3.1  &1.67  &1.  \\
0.2235  &-19.76  &0.62  &1.95  &1.  &\nodata &12.7  &\nodata &0.60  &1.7  &-7.6  &2.39  &0.  \\
0.2155  &-18.75  &0.63  &1.20  &0.  &21.4  &\nodata &0.72  &\nodata &-44.4  &4.3  &1.61  &1.  \\
0.2324  &-18.94  &0.63  &1.68  &0.36  &20.8  &14.6  &0.51  &0.53  &6.9  &3.4  &1.96  &0.  \\
0.2334  &-18.51  &0.64  &1.83  &0.  &20.0  &\nodata &0.36  &\nodata &-12.6  &6.1  &2.19  &\nodata  \\
0.2253  &-19.63  &0.64  &1.45  &0.  &20.0  &\nodata &0.60  &\nodata &-5.8  &6.9  &1.81  &\nodata  \\
0.2351  &-19.85  &0.66  &1.80  &0.42  &21.3  &11.7  &0.74  &0.13  &7.8  &1.1  &2.53  &\nodata  \\
0.2323  &-19.02  &0.68  &1.84  &1.  &\nodata &11.9  &\nodata &0.47  &0.9  &2.7  &2.36  &3.  \\
0.2278  &-18.53  &0.69  &1.17  &0.  &20.2  &\nodata &0.42  &\nodata &-12.7  &5.3  &1.76  &\nodata  \\
0.2257  &-18.74  &0.71  &1.90  &1.  &\nodata &12.4  &\nodata &0.55  &2.6  &2.0  &2.00  &\nodata  \\
0.2326  &-17.82  &0.71  &1.56  &0.  &20.7  &\nodata &0.37  &\nodata &-9.9  &6.9  &2.10  &\nodata  \\
0.2240  &-20.09  &0.72  &1.83  &0.57  &21.0  &14.2  &0.68  &0.77  &3.0  &-1.0  &2.69  &1.  \\
0.2230  &-18.46  &0.72  &1.91  &1.  &\nodata &11.2  &\nodata &-0.02  &-22.3  &-0.2  &2.71  &\nodata  \\
0.2279  &-19.25  &0.73  &1.60  &1.  &\nodata &13.5  &\nodata &0.65  &-2.5  &3.2  &1.96  &0.  \\
0.2312  &-19.16  &0.73  &1.90  &0.43  &22.3  &12.1  &0.83  &0.13  &21.4  &1.2  &2.27  &\nodata  \\
0.2239  &-18.73  &0.76  &1.96  &0.31  &20.8  &10.0  &0.44  &-0.50  &75.2  &-4.9  &2.70  &\nodata  \\
0.2265  &-17.60  &0.77  &1.75  &0.  &20.0  &\nodata &0.14  &\nodata &5.7  &3.6  &2.32  &\nodata  \\
0.2248  &-19.15  &0.77  &1.10  &0.  &19.9  &\nodata &0.50  &\nodata &5.2  &2.8  &1.63  &0.  \\
0.2279  &-17.21  &0.77  &0.74  &0.  &21.0  &\nodata &0.37  &\nodata &-21.2  &12.  &1.64  &\nodata  \\
0.2254  &-18.66  &0.78  &1.78  &1.  &\nodata &13.7  &\nodata &0.63  &-12.8  &2.9  &1.74  &\nodata  \\
0.2183  &-18.21  &0.80  &1.66  &1.  &\nodata &13.3  &\nodata &0.60  &-9.6  &1.8  &2.47  &\nodata  \\
0.2341  &-19.33  &0.81  &1.84  &1.  &\nodata &12.7  &\nodata &0.58  &4.9  &-2.6  &2.39  &\nodata  \\
0.2331  &-17.84  &0.82  &1.86  &1.  &\nodata &13.2  &\nodata &0.27  &45.2  &4.6  &1.96  &\nodata  \\
0.2257  &-17.41  &0.82  &1.18  &0.  &21.4  &\nodata &0.49  &\nodata &-26.1  &12.1  &1.57  &\nodata  \\
0.2228  &-19.19  &0.84  &1.64  &1.  &\nodata &12.9  &\nodata &0.50  &-9.1  &-2.7  &1.76  &0.  \\
0.2163  &-19.55  &0.86  &1.76  &0.53  &20.6  &12.0  &0.51  &0.20  &9.0  &3.9  &2.41  &0.  \\
0.2332  &-18.72  &0.87  &1.77  &1.  &\nodata &14.8  &\nodata &0.87  &6.0  &4.4  &2.55  &2.  \\
0.2469  &-19.25  &0.90  &1.79  &1.  &\nodata &14.1  &\nodata &0.79  &6.3  &1.0  &2.77  &2.  \\
0.2271  &-18.82  &0.90  &1.58  &0.61  &20.2  &15.1  &0.30  &0.76  &-13.6  &0.5  &1.95  &\nodata  \\
0.2371  &-18.49  &0.90  &1.63  &0.05  &21.5  &10.3  &0.65  &-0.84  &-3.7  &5.6  &2.15  &2.  \\
0.2302  &-17.75  &0.90  &1.71  &0.  &20.0  &\nodata &0.17  &\nodata &3.8  &4.9  &2.38  &3.  \\
0.2284  &-17.26  &0.91  &1.59  &0.  &20.9  &\nodata &0.28  &\nodata &4.9  &0.3  &2.21  &1.  \\
0.2199  &-18.73  &0.94  &1.78  &0.  &20.7  &\nodata &0.52  &\nodata &1.6  &3.6  &2.43  &\nodata  \\
0.2286  &-19.49  &0.94  &1.82  &0.23  &21.5  &15.9  &0.78  &0.79  &8.0  &2.6  &2.34  &\nodata  \\
0.2329  &-19.63  &0.95  &1.40  &0.12  &20.1  &8.2  &0.57  &-0.84  &-8.5  &2.5  &1.65  &\nodata  \\
0.2282  &-19.58  &0.95  &2.08  &0.25  &22.5  &12.8  &1.06  &0.27  &-5.9  &15.7  &2.38  &3.  \\
0.2348  &-19.41  &0.96  &1.84  &1.  &\nodata &14.5  &\nodata &0.88  &-6.0  &4.8  &2.44  &1.  \\
0.2249  &-19.79  &0.97  &1.85  &0.59  &22.0  &10.9  &0.80  &0.05  &2.8  &-2.2  &2.58  &\nodata  \\
0.2284  &-18.83  &0.97  &1.84  &1.  &\nodata &13.8  &\nodata &0.78  &6.2  &4.2  &2.23  &0.  \\
0.2363  &-18.24  &0.97  &1.70  &0.  &19.9  &\nodata &0.27  &\nodata &4.7  &3.9  &2.10  &\nodata  \\
0.2265  &-19.67  &0.99  &1.89  &0.65  &23.1  &13.0  &0.96  &0.45  &-3.3  &3.5  &2.75  &\nodata  \\
0.2269  &-18.80  &0.99  &1.71  &0.39  &20.5  &9.5  &0.39  &-0.53  &6.8  &4.0  &2.27  &\nodata  \\
0.2233  &-19.23  &1.00  &1.87  &0.08  &21.0  &9.2  &0.67  &-0.84  &-16.9  &-1.9  &2.46  &1.  \\
0.2251  &-19.47  &1.05  &1.75  &0.06  &21.4  &9.3  &0.81  &-0.82  &-1.2  &6.0  &2.22  &0.  \\
0.2274  &-17.65  &1.06  &1.73  &1.  &\nodata &15.1  &\nodata &0.74  &-16.4  &-0.3  &2.10  &1.  \\
0.2353  &-20.24  &1.07  &1.86  &0.39  &21.4  &11.1  &0.83  &0.06  &1.3  &-1.7  &2.45  &2.  \\
0.2244  &-18.43  &1.13  &1.53  &0.61  &21.0  &13.9  &0.35  &0.40  &7.3  &9.6  &2.12  &3.  \\
0.2287  &-19.40  &1.13  &1.58  &0.18  &20.2  &8.2  &0.51  &-0.47  &-9.2  &4.4  &2.13  &0.  \\
0.2315  &-19.30  &1.14  &1.60  &0.18  &20.1  &14.5  &0.50  &0.45  &3.0  &3.9  &1.88  &1.  \\
0.2226  &-19.41  &1.15  &1.88  &0.13  &20.9  &9.5  &0.74  &-0.56  &-1.2  &1.2  &2.20  &1.  \\
0.2354  &-19.12  &1.15  &1.78  &1.  &\nodata &12.2  &\nodata &0.62  &3.1  &0.0  &2.61  &\nodata  \\
0.2301  &-19.33  &1.16  &1.86  &0.39  &21.1  &7.4  &0.60  &-0.84  &4.7  &-1.2  &2.81  &\nodata  \\
0.2345  &-18.33  &1.17  &1.78  &0.63  &20.9  &13.5  &0.28  &0.31  &2.1  &1.0  &2.06  &4.  \\
0.2544  &-19.66  &1.18  &1.88  &0.53  &23.0  &11.4  &0.99  &0.10  &-5.5  &3.7  &2.31  &0.  \\
0.2290  &-19.23  &1.20  &1.37  &1.  &\nodata &14.1  &\nodata &0.85  &0.8  &0.5  &\nodata  &2.  \\
0.2248  &-18.03  &1.20  &1.59  &1.  &\nodata &14.0  &\nodata &0.80  &-50.3  &8.1  &2.60  &\nodata  \\
0.2599  &-18.22  &1.27  &1.61  &1.  &\nodata &12.8  &\nodata &0.26  &9.8  &5.4  &2.27  &1.  \\
0.2254  &-18.50  &1.29  &1.53  &0.  &20.2  &\nodata &0.38  &\nodata &-8.6  &4.1  &2.08  &0.  \\
0.2265  &-18.72  &1.31  &1.78  &1.  &\nodata &11.7  &\nodata &0.28  &6.9  &1.6  &2.16  &0.  \\
0.2291  &-19.38  &1.35  &1.53  &0.  &20.8  &\nodata &0.67  &\nodata &-7.2  &3.7  &2.13  &\nodata  \\
0.2267  &-18.86  &1.35  &1.84  &1.  &\nodata &12.9  &\nodata &0.60  &7.9  &-0.4  &2.33  &1.  \\
0.2357  &-18.40  &1.35  &1.62  &0.  &20.5  &\nodata &0.43  &\nodata &-45.5  &5.8  &1.83  &1.  \\
0.2232  &-19.91  &1.42  &1.84  &1.  &\nodata &14.1  &\nodata &0.96  &6.6  &1.3  &2.39  &\nodata  \\
0.2287  &-19.05  &1.47  &1.75  &1.  &\nodata &12.6  &\nodata &0.52  &4.5  &0.2  &2.63  &0.  \\
0.2228  &-19.65  &1.50  &1.83  &0.42  &21.3  &10.6  &0.70  &-0.12  &-2.0  &1.7  &2.70  &4.  \\
0.2333  &-18.85  &1.58  &1.77  &1.  &\nodata &12.9  &\nodata &0.64  &0.3  &3.1  &2.64  &\nodata  \\
0.2221  &-18.99  &1.63  &1.60  &1.  &\nodata &13.4  &\nodata &0.63  &-1.2  &6.8  &1.71  &\nodata  \\
0.2227  &-17.74  &1.70  &1.56  &1.  &\nodata &13.5  &\nodata &0.35  &4.0  &1.8  &1.98  &1.  \\
0.2232  &-18.91  &1.79  &0.75  &0.  &19.6  &\nodata &0.40  &\nodata &-16.1  &2.4  &1.46  &\nodata  \\
0.2133  &-18.09  &1.87  &1.69  &1.  &\nodata &13.5  &\nodata &0.41  &5.5  &4.0  &2.03  &4.  \\
0.2329  &-18.78  &1.88  &1.16  &0.13  &20.6  &12.9  &0.53  &-0.04  &8.0  &8.4  &2.07  &2.  \\
0.2252  &-18.80  &2.10 &2.01  &0.09  &20.7  &11.1  &0.50  &-0.54  &-22.7  &4.1  &2.22  &1.  \\

\enddata
\end{deluxetable}

\begin{deluxetable}{lllclrrrrrrc}
\scriptsize
\tablecaption{Field Galaxies}
\tablehead{
\colhead{z} & \colhead{$M_{B}$}  & \colhead{(U-V)$_{0}$} 
&\colhead{$B/T$} & \colhead{$\mu_{disk}$} & \colhead{$\mu_{bulge}$} 
&\colhead{Scl.ht}
&\colhead{1/2lt rad} &\colhead{$[O II]$} & \colhead{$H_{\delta}$} & \colhead{D4000} 
& \colhead{Int}\\
&\colhead{} &&&&&\colhead{log(kpc)} &\colhead{log(kpc)}
&\colhead{(\AA)} &\colhead{(\AA)} &\colhead{} &\colhead{level}\\ }
\startdata

0.0658  &-17.77  &0.84  &0.  &18.52  &\nodata &-0.1  &\nodata &\nodata  &\nodata  &\nodata  &0.  \\
0.0669  &-14.34  &0.59  &0.47  &20.51  &10.21  &-0.6  &-1.3  &\nodata  &6.5  &\nodata  &0.  \\
0.0683  &-14.83  &1.50  &0.  &20.76  &\nodata &-0.3  &\nodata &\nodata  &9.8  &\nodata  &\nodata  \\
0.1029  &-17.80  &1.24  &0.  &20.83  &\nodata &0.3  &\nodata &\nodata  &16.4  &\nodata  &\nodata  \\
0.1149  &-20.06  &2.14  &1.  &\nodata &13.45  &\nodata &0.7  &\nodata  &2.0  &\nodata  &\nodata  \\
0.1264  &-18.00  &2.45  &0.47  &21.77  &10.46  &0.3  &-0.6  &\nodata  &-17.7  &\nodata  &0.  \\
0.1472  &-17.76  &1.30  &0.  &20.65  &\nodata &0.3  &\nodata &\nodata  &6.6  &1.47  &\nodata  \\
0.1588  &-19.46  &1.37  &0.37  &20.66  &14.06  &0.5  &0.5  &\nodata  &5.7  &1.74  &4.  \\
0.1689  &-18.03  &1.79  &1.  &\nodata &10.91  &\nodata &0.2  &\nodata  &1.3  &2.11  &\nodata  \\
0.1699  &-20.72  &1.57  &1.  &\nodata &16.12  &\nodata &1.6  &\nodata  &6.9  &1.83  &\nodata  \\
0.1740  &-18.71  &1.39  &0.  &20.24  &\nodata &0.4  &\nodata &\nodata  &6.2  &1.70  &\nodata  \\
0.1755  &-18.51  &1.97  &0.31  &20.27  &11.13  &0.3  &-0.3  &\nodata  &0.5  &2.19  &\nodata  \\
0.1783  &-16.90  &1.70  &0.  &19.46  &\nodata &-0.1  &\nodata &12.3  &7.8  &1.49  &1.  \\
0.1785  &-20.57  &2.00  &0.43  &21.92  &12.71  &1.0  &0.5  &1.8  &-0.4  &2.41  &\nodata  \\
0.1794  &-20.05  &1.58  &0.2  &19.84  &15.06  &0.5  &0.7  &-36.  &2.8  &2.64  &\nodata  \\
0.1796  &-19.42  &2.05  &0.31  &20.85  &9.99  &0.6  &-0.4  &-3.  &1.2  &2.42  &\nodata  \\
0.1799  &-19.25  &2.17  &0.07  &21.6  &11.19  &0.7  &-0.5  &17.2  &11.2  &1.57  &\nodata  \\
0.1800  &-17.63  &1.70  &0.  &19.65  &\nodata &0.1  &\nodata &-20.3  &2.3  &1.88  &\nodata  \\
0.1851  &-18.45  &1.91  &0.31  &21.34  &8.81  &0.5  &-0.8  &-29.7  &0.0  &2.52  &\nodata  \\
0.1861  &-18.05  &1.80  &0.  &19.84  &\nodata &0.2  &\nodata &-8.5  &7.2  &1.78  &\nodata  \\
0.1879  &-17.95  &0.98  &0.  &20.36  &\nodata &0.3  &\nodata &-38.3  &4.4  &1.77  &1.  \\
0.1898  &-19.08  &2.19  &0.77  &21.12  &14.94  &0.3  &0.7  &\nodata  &\nodata  &-0.42  &2.  \\
0.2028  &-18.18  &1.56  &0.  &19.93  &\nodata &0.2  &\nodata &-22.5  &4.5  &1.83  &\nodata  \\
0.2032  &-18.12  &1.26  &1.  &\nodata &12.65  &\nodata &0.3  &-13.6  &5.5  &1.88  &\nodata  \\
0.2047  &-19.50  &1.12  &0.  &20.04  &\nodata &0.6  &\nodata &-9.3  &12.6  &1.90  &\nodata  \\
0.2060  &-18.81  &1.52  &0.03  &21.75  &10.27  &0.8  &-0.9  &-24.5  &6.5  &1.97  &\nodata  \\
0.2544  &-18.52  &1.61  &1.  &\nodata &11.68  &\nodata &0.2  &3.3  &0.5  &2.30  &0.  \\
0.2551  &-20.30  &1.55  &0.76  &19.84  &14.77  &0.4  &1.  &-3.7  &6.9  &2.01  &\nodata  \\
0.2579  &-17.95  &0.71  &0.  &21.3  &\nodata &0.5  &\nodata &-51.3  &4.0  &1.48  &\nodata  \\
0.2599  &-18.54  &0.69  &0.  &20.54  &\nodata &0.5  &\nodata &-7.5  &5.9  &1.66  &1.  \\
0.2617  &-17.54  &2.15  &0.14  &22.62  &16.17  &0.7  &0.4  &-13.4  &-10.8  &1.34  &\nodata  \\
0.2646  &-17.45  &0.73  &0.  &20.01  &\nodata &0.2  &\nodata &-46.2  &5.1  &1.43  &\nodata  \\
0.2654  &-18.28  &0.97  &0.  &21.24  &\nodata &0.6  &\nodata &-14.1  &1.9  &1.91  &\nodata  \\
0.2661  &-16.92  &0.95  &0.  &19.04  &\nodata &-0.1  &\nodata &-74.7  &6.6  &1.86  &0.  \\
0.2741  &-18.86  &1.32  &0.  &19.65  &\nodata &0.4  &\nodata &-1.6  &6.5  &1.83  &0.  \\
0.2745  &-18.51  &1.52  &1.  &\nodata &20.21  &\nodata &2.  &-4.1  &0.7  &1.96  &\nodata  \\
0.2751  &-16.63  &0.63  &0.  &24.76  &\nodata &1.5  &\nodata &-63.3  &9.4  &1.14  &\nodata  \\
0.2797  &-19.75  &1.42  &0.1  &21.08  &8.45  &0.8  &-0.8  &-7.5  &4.1  &2.11  &0.  \\
0.2810  &-20.16  &1.64  &0.25  &21.48  &11.76  &0.9  &0.2  &-6.8  &-0.4  &1.97  &1.  \\
0.2819  &-19.78  &1.48  &1.  &\nodata &14.94  &\nodata &1.2  &-5.6  &3.5  &1.81  &1.  \\
0.2827  &-16.85  &1.00  &0.  &20.  &\nodata &0.1  &\nodata &-43.7  &7.9  &2.03  &\nodata  \\
0.2834  &-18.43  &1.10  &0.  &20.93  &\nodata &0.6  &\nodata &-9.  &8.1  &1.69  &0.  \\
0.3042  &-17.53  &1.36  &0.89  &22.8  &9.67  &0.3  &-0.5  &-9.  &-1.0  &1.91  &\nodata  \\
0.3066  &-20.06  &1.75  &0.09  &21.02  &8.32  &0.9  &-0.7  &-7.5  &-1.4  &1.92  &\nodata  \\
0.3068  &-17.53  &0.65  &0.  &20.33  &\nodata &0.7  &\nodata &-28.3  &11.1  &1.97  &\nodata  \\
0.3069  &-17.94  &1.28  &0.77  &27.52  &11.64  &1.5  &0.  &-22.5  &4.1  &1.76  &\nodata  \\
0.3071  &-19.58  &1.05  &0.  &20.47  &\nodata &0.7  &\nodata &-11.6  &2.0  &2.09  &2.  \\
0.3076  &-17.83  &0.92  &0.  &20.99  &\nodata &0.5  &\nodata &-34.1  &4.7  &2.09  &3.  \\
0.3077  &-19.71  &1.17  &0.22  &22.24  &14.15  &1.1  &0.6  &-8.0  &1.0  &1.77  &\nodata  \\
0.3080  &-18.66  &1.17  &0.  &19.99  &\nodata &0.5  &\nodata &-26.8  &3.3  &1.75  &1.  \\
0.3093  &-17.75  &0.60  &0.  &19.31  &\nodata &0.1  &\nodata &-40.1  &5.3  &1.39  &\nodata  \\
0.3093  &-17.26  &0.54  &0.  &21.08  &\nodata &0.4  &\nodata &-38.4  &10.3  &1.41  &1.  \\
0.3097  &-17.74  &1.79  &0.22  &25.81  &9.81  &1.6  &-0.5  &5.2  &0.4  &1.93  &\nodata  \\
0.3125  &-17.82  &0.67  &1.  &\nodata &14.53  &\nodata &0.9  &-33.6  &9.5  &1.35  &\nodata  \\
0.3149  &-18.23  &1.04  &0.  &20.4  &\nodata &0.4  &\nodata &-21.4  &12.  &1.79  &\nodata  \\
0.3167  &-20.28  &1.87  &0.37  &21.32  &11.6  &0.9  &0.2  &-12.9  &-3.6  &2.86  &\nodata  \\
0.3179  &-20.04  &1.63  &1.  &\nodata &14.44  &\nodata &1.2  &-11.6  &-0.7  &2.15  &\nodata  \\
0.3199  &-20.11  &1.86  &0.63  &22.78  &12.39  &1.1  &0.5  &-4.1  &-0.4  &\nodata  &\nodata  \\
0.3203  &-18.87  &1.19  &0.  &20.3  &\nodata &0.6  &\nodata &-15.  &9.3  &\nodata  &3.  \\
0.3207  &-17.12  &0.75  &0.51  &24.58  &12.48  &1.0  &0.  &-30.3  &12.1  &\nodata  &\nodata  \\
0.3207  &-16.75  &1.01  &0.  &22.86  &\nodata &0.7  &\nodata &-65.1  &11.6  &\nodata  &1.  \\
0.3209  &-18.11  &0.76  &0.  &19.74  &\nodata &0.3  &\nodata &-20.3  &3.6  &\nodata  &\nodata  \\
0.3217  &-19.84  &1.21  &0.27  &20.17  &12.44  &0.7  &0.3  &-6.3  &4.3  &\nodata  &\nodata  \\
0.3221  &-18.15  &1.73  &0.  &20.18  &\nodata &0.4  &\nodata &-2.6  &0.9  &\nodata  &\nodata  \\
0.3351  &-19.60  &1.40  &1.  &\nodata &10.44  &\nodata &0.4  &0.0  &1.5  &\nodata  &1.  \\
0.3396  &-19.29  &1.66  &1.  &\nodata &12.97  &\nodata &0.8  &-5.9  &-5.7  &\nodata  &3.  \\
0.3399  &-18.61  &1.35  &1.  &\nodata &12.14  &\nodata &0.4  &-6.4  &5.4  &\nodata  &0.  \\
0.3471  &-17.69  &1.26  &0.  &19.01  &\nodata &0.1  &\nodata &-22.  &\nodata  &\nodata  &0.  \\
0.3474  &-18.14  &1.38  &0.14  &25.23  &12.26  &1.7  &0.1  &6.6  &\nodata  &\nodata  &\nodata  \\
0.3474  &-19.57  &1.67  &0.62  &23.21  &13.27  &1.1  &0.6  &-1.4  &\nodata  &\nodata  &\nodata  \\
0.3475  &-18.65  &0.82  &0.  &20.69  &\nodata &0.6  &\nodata &2.8  &\nodata  &\nodata  &\nodata  \\
0.3476  &-17.71  &0.89  &1.  &\nodata &11.76  &\nodata &0.2  &-35.8  &\nodata  &\nodata  &\nodata  \\
0.3477  &-19.44  &1.62  &0.25  &23.08  &12.63  &1.4  &0.4  &10.1  &\nodata  &\nodata  &\nodata  \\
0.3479  &-18.67  &1.73  &0.  &16.63  &\nodata &0.6  &\nodata &5.5  &\nodata  &\nodata  &\nodata  \\
0.3484  &-18.42  &1.33  &0.  &20.37  &\nodata &0.5  &\nodata &-13.4  &\nodata  &\nodata  &\nodata  \\
0.3484  &-19.18  &1.37  &0.48  &20.75  &15.77  &0.6  &1.  &-3.6  &\nodata  &\nodata  &\nodata  \\
0.3486  &-17.49  &0.59  &0.  &20.99  &\nodata &0.4  &\nodata &-40.9  &\nodata  &\nodata  &\nodata  \\
0.3486  &-19.25  &1.09  &1.  &\nodata &14.12  &\nodata &0.9  &-17.0  &\nodata  &\nodata  &\nodata  \\
0.3488  &-20.22  &1.73  &0.51  &21.28  &12.18  &0.9  &0.4  &-1.1  &\nodata  &\nodata  &\nodata  \\
0.3546  &-18.53  &1.31  &0.  &21.09  &\nodata &0.7  &\nodata &-9.4  &\nodata  &\nodata  &\nodata  \\
0.3697  &-18.17  &0.87  &0.  &21.52  &\nodata &0.7  &\nodata &-37.2  &\nodata  &\nodata  &\nodata  \\
0.3829  &-18.30  &1.01  &0.  &19.65  &\nodata &0.4  &\nodata &-25.8  &\nodata  &\nodata  &\nodata  \\
0.3830  &-17.88  &1.34  &0.  &20.08  &\nodata &0.3  &\nodata &-24.7  &\nodata  &\nodata  &0.  \\
0.3834  &-18.24  &0.90  &0.35  &26.48  &11.96  &1.8  &0.2  &-30.4  &\nodata  &\nodata  &0.  \\
0.3885  &-19.99  &0.20  &0.  &21.46  &\nodata &1.1  &\nodata &-16.1  &\nodata  &\nodata  &\nodata  \\
0.3896  &-19.99  &0.35  &0.18  &19.71  &13.88  &0.7  &0.6  &-12.2  &\nodata  &\nodata  &0.  \\
0.3978  &-18.38  &1.10  &0.  &20.05  &\nodata &0.4  &\nodata &-17.4  &\nodata  &\nodata  &\nodata  \\
0.3981  &-18.45  &1.41  &1.  &\nodata &12.56  &\nodata &0.5  &-16.4  &\nodata  &\nodata  &\nodata  \\
0.3984  &-17.95  &0.81  &0.  &20.43  &\nodata &0.4  &\nodata &-28.3  &\nodata  &\nodata  &1.  \\
0.4032  &-17.89  &0.04  &1.  &\nodata &12.17  &\nodata &0.5  &9.8  &\nodata  &\nodata  &1.  \\
0.4091  &-17.84  &1.08  &0.  &21.47  &\nodata &0.6  &\nodata &-51.5  &\nodata  &\nodata  &1.  \\
0.4115  &-19.45  &1.22  &0.  &18.88  &\nodata &0.4  &\nodata &-17.0  &\nodata  &\nodata  &\nodata  \\
0.4115  &-18.68  &1.01  &0.  &19.83  &\nodata &0.5  &\nodata &-12.5  &\nodata  &\nodata  &0.  \\
0.4274  &-17.05  &0.32  &0.01  &25.92  &9.06  &2.3  &-0.7  &6.5  &\nodata  &\nodata  &\nodata  \\
0.4349  &-18.09  &1.39  &1.  &\nodata &15.16  &\nodata &1.3  &-14.7  &\nodata  &\nodata  &\nodata  \\
0.4359  &-19.88  &1.38  &0.22  &18.29  &7.49  &0.4  &-0.7  &-2.9  &\nodata  &\nodata  &0.  \\
0.4548  &-18.73  &1.11  &0.  &19.57  &\nodata &0.4  &\nodata &-23.5  &\nodata  &\nodata  &3.  \\
0.2099  &-17.52  &0.83  &0.  &21.08  &\nodata &0.1  &\nodata &\nodata  &1.5  &\nodata  &1.  \\
0.2067  &-17.08  &0.91  &0.  &22.56  &\nodata &0.5  &\nodata &\nodata  &4.3  &\nodata  &1.  \\
0.2061  &-17.55  &0.78  &0.  &21.87  &\nodata &0.5  &\nodata &\nodata  &2.2  &\nodata  &2.  \\
0.2113  &-17.50  &0.88  &0.  &20.3  &\nodata &0.2  &\nodata &\nodata  &8.8  &\nodata  &0.  \\
0.2462  &-17.42  &0.97  &0.  &22.03  &\nodata &0.5  &\nodata &-47.9  &-2.9  &1.37  &1.  \\
0.2656  &-17.43  &1.91  &0.  &21.68  &\nodata &0.5  &\nodata &\nodata  &2.6  &2.02  &1.  \\
0.2064  &-17.57  &0.95  &0.  &20.91  &\nodata &0.3  &\nodata &\nodata  &8.9  &\nodata  &2.  \\
0.2516  &-17.74  &0.82  &0.  &20.8  &\nodata &0.3  &\nodata &\nodata  &8.2  &1.68  &1.  \\
0.2210  &-17.66  &2.05  &0.02  &21.13  &13.95  &0.4  &-0.5  &\nodata  &-0.2  &\nodata  &0.  \\
0.2065  &-17.67  &0.71  &0.  &21.43  &\nodata &0.4  &\nodata &\nodata  &1.1  &\nodata  &0.  \\
0.2428  &-17.60  &0.53  &0.  &21.96  &\nodata &0.6  &\nodata &\nodata  &4.2  &1.09  &0.  \\
0.2221  &-18.12  &1.92  &0.79  &22.25  &9.95  &0.3  &-0.5  &\nodata  &7.0  &\nodata  &0.  \\
0.2583  &-17.78  &1.97  &0.  &20.28  &\nodata &0.2  &\nodata &\nodata  &2.1  &1.93  &0.  \\
0.2076  &-17.77  &1.26  &0.06  &20.  &12.43  &0.2  &-0.6  &-37.5  &10.6  &1.14  &0.  \\
0.2090  &-17.86  &1.48  &0.  &20.65  &\nodata &0.3  &\nodata &-60.9  &1.7  &1.61  &0.  \\
0.2079  &-17.30  &1.63  &0.47  &21.64  &19.59  &0.4  &1.3  &-111.4  &1.2  &1.22  &0.  \\
0.2561  &-17.76  &1.65  &0.  &21.84  &\nodata &0.6  &\nodata &\nodata  &1.9  &1.32  &3.  \\
0.2673  &-17.93  &0.45  &0.  &21.15  &\nodata &0.4  &\nodata &\nodata  &1.3  &1.02  &0.  \\
0.2315  &-17.85  &1.47  &0.  &19.38  &\nodata &0.1  &\nodata &-26.4  &-0.7  &1.75  &\nodata  \\
0.2646  &-17.45  &0.73  &0.  &20.01  &\nodata &0.2  &\nodata &-46.2  &5.1  &1.43  &0.  \\
0.1948  &-17.88  &1.14  &0.  &19.23  &\nodata &0.1  &\nodata &\nodata  &5.4  &\nodata  &1.  \\
0.2103  &-18.10  &1.01  &0.62  &22.36  &9.66  &0.5  &-0.6  &\nodata  &2.1  &\nodata  &0.  \\
0.2045  &-18.31  &1.07  &0.  &20.01  &\nodata &0.2  &\nodata &\nodata  &6.4  &\nodata  &2.  \\
0.2669  &-17.81  &1.37  &0.  &21.43  &\nodata &0.5  &\nodata &\nodata  &7.7  &1.52  &0.  \\
0.2674  &-18.09  &1.07  &0.  &20.21  &\nodata &0.3  &\nodata &\nodata  &3.5  &1.43  &0.  \\
0.2435  &-18.02  &1.35  &0.  &20.26  &\nodata &0.3  &\nodata &-43.9  &8.9  &1.67  &0.  \\
0.2028  &-18.18  &1.56  &0.  &19.93  &\nodata &0.2  &\nodata &-22.5  &4.5  &1.83  &0.  \\
0.2380  &-18.10  &1.93  &1.  &\nodata &12.22  &\nodata &0.3  &\nodata  &11.8  &\nodata  &0.  \\
0.2655  &-17.85  &1.71  &1.  &\nodata &14.33  &\nodata &0.5  &\nodata  &1.3  &1.84  &1.  \\
0.2064  &-18.34  &1.31  &0.  &21.24  &\nodata &0.5  &\nodata &\nodata  &4.2  &\nodata  &3.  \\
0.2579  &-17.95  &0.71  &0.  &21.3  &\nodata &0.5  &\nodata &-51.3  &4.0  &1.48  &1.  \\
0.2210  &-18.42  &1.98  &1.  &\nodata &11.09  &\nodata &0.1  &\nodata  &2.2  &\nodata  &1.  \\
0.2032  &-18.12  &1.26  &1.  &\nodata &12.65  &\nodata &0.3  &-13.6  &5.5  &1.88  &1.  \\
0.2442  &-17.89  &0.48  &0.  &20.23  &\nodata &0.3  &\nodata &-76.2  &15.4  &1.43  &2.  \\
0.2656  &-18.42  &1.90  &0.  &21.5  &\nodata &0.6  &\nodata &\nodata  &5.4  &1.62  &2.  \\
0.2301  &-18.45  &0.88  &0.  &21.04  &\nodata &0.5  &\nodata &\nodata  &7.9  &\nodata  &1.  \\
0.2237  &-18.66  &1.87  &0.43  &21.34  &9.77  &0.5  &-0.5  &\nodata  &7.8  &\nodata  &3.  \\
0.2390  &-18.43  &0.58  &0.  &21.49  &\nodata &0.6  &\nodata &\nodata  &1.8  &\nodata  &1.  \\
0.2260  &-18.57  &1.04  &0.22  &21.16  &15.88  &0.5  &0.5  &\nodata  &5.9  &\nodata  &0.  \\
0.2250  &-18.74  &2.15  &0.41  &22.33  &9.78  &0.7  &-0.5  &\nodata  &3.1  &\nodata  &2.  \\
0.2109  &-18.56  &1.32  &0.  &19.35  &\nodata &0.2  &\nodata &\nodata  &6.3  &\nodata  &0.  \\
0.2206  &-18.60  &1.92  &1.  &\nodata &12.59  &\nodata &0.6  &-14.4  &2.2  &1.82  &3.  \\
0.2255  &-18.51  &1.62  &0.42  &20.47  &9.71  &0.3  &-0.5  &\nodata  &3.3  &\nodata  &0.  \\
0.2509  &-18.54  &1.44  &0.21  &19.68  &10.58  &0.2  &-0.5  &\nodata  &1.3  &1.73  &0.  \\
0.2545  &-18.54  &1.71  &0.  &20.85  &\nodata &0.5  &\nodata &\nodata  &2.2  &1.39  &0.  \\
0.2083  &-18.71  &1.69  &0.29  &21.96  &14.01  &0.7  &0.3  &0.9  &3.5  &1.44  &1.  \\
0.2654  &-18.28  &0.97  &0.  &21.24  &\nodata &0.6  &\nodata &-14.1  &1.9  &1.91  &1.  \\
0.2096  &-18.76  &1.77  &1.  &\nodata &12.35  &\nodata &0.3  &\nodata  &7.4  &\nodata  &1.  \\
0.2304  &-18.64  &1.25  &0.  &20.14  &\nodata &0.4  &\nodata &\nodata  &6.2  &\nodata  &0.  \\
0.2685  &-18.51  &0.81  &0.  &19.46  &\nodata &0.2  &\nodata &\nodata  &3.6  &1.61  &0.  \\
0.1928  &-18.75  &0.89  &0.  &21.18  &\nodata &0.6  &\nodata &\nodata  &8.2  &\nodata  &1.  \\
0.2068  &-18.68  &1.03  &0.  &22.52  &\nodata &0.9  &\nodata &\nodata  &6.5  &\nodata  &1.  \\
0.2244  &-18.62  &1.50  &0.53  &21.16  &12.67  &0.4  &0.1  &\nodata  &2.3  &\nodata  &0.  \\
0.2495  &-18.80  &1.70  &1.  &\nodata &11.7  &\nodata &0.3  &\nodata  &\nodata  &\nodata  &1.  \\
0.2000  &-18.67  &0.64  &0.  &20.79  &\nodata &0.5  &\nodata &\nodata  &11.5  &\nodata  &4.  \\
0.2250  &-18.71  &1.45  &0.33  &19.82  &10.49  &0.2  &-0.4  &\nodata  &7.2  &\nodata  &0.  \\
0.2461  &-18.84  &0.82  &0.  &20.4  &\nodata &0.4  &\nodata &\nodata  &6.3  &1.82  &0.  \\
0.2544  &-18.52  &1.61  &1.  &\nodata &11.68  &\nodata &0.2  &3.3  &0.5  &2.30  &0.  \\
0.2213  &-18.82  &1.74  &0.18  &20.42  &11.64  &0.4  &-0.3  &\nodata  &4.9  &\nodata  &0.  \\
0.2597  &-18.64  &1.53  &1.  &\nodata &12.33  &\nodata &0.4  &\nodata  &2.3  &2.62  &0.  \\
0.1912  &-18.98  &2.00  &0.66  &22.69  &14.63  &0.7  &0.6  &\nodata  &-0.6  &\nodata  &0.  \\
0.2069  &-19.02  &2.06  &1.  &\nodata &11.75  &\nodata &0.4  &8.5  &3.6  &2.97  &0.  \\
0.2126  &-18.71  &1.64  &0.27  &20.32  &19.44  &0.4  &1.4  &-0.5  &9.4  &1.81  &1.  \\
0.2469  &-18.78  &0.98  &0.  &20.95  &\nodata &0.6  &\nodata &-16.6  &8.4  &1.65  &0.  \\
0.2446  &-18.71  &1.25  &0.13  &21.25  &15.69  &0.6  &0.5  &\nodata  &3.9  &1.43  &4.  \\
0.2235  &-18.97  &1.01  &0.  &21.21  &\nodata &0.7  &\nodata &\nodata  &4.3  &\nodata  &1.  \\
0.2247  &-19.09  &2.06  &0.38  &21.63  &10.3  &0.6  &-0.4  &\nodata  &-0.4  &\nodata  &1.  \\
0.2059  &-19.06  &1.72  &0.58  &19.83  &13.2  &0.2  &0.3  &\nodata  &0.2  &\nodata  &0.  \\
0.2114  &-19.05  &1.85  &0.37  &21.03  &9.3  &0.5  &-0.6  &\nodata  &0.4  &\nodata  &1.  \\
0.2599  &-18.54  &0.69  &0.  &20.54  &\nodata &0.5  &\nodata &-7.5  &5.9  &1.66  &0.  \\
0.2485  &-18.91  &1.60  &0.29  &20.56  &9.76  &0.5  &-0.5  &\nodata  &2.  &2.41  &0.  \\
0.2163  &-18.57  &0.87  &0.05  &21.35  &10.04  &0.7  &-0.9  &-15.8  &2.7  &1.48  &1.  \\
0.2060  &-18.81  &1.52  &0.03  &21.75  &10.27  &0.8  &-0.9  &-24.5  &6.5  &1.97  &0.  \\
0.2656  &-19.21  &1.91  &0.55  &21.37  &12.11  &0.5  &0.1  &-15.6  &3.8  &2.00  &0.  \\
0.2215  &-19.49  &2.09  &0.  &19.77  &\nodata &0.4  &\nodata &\nodata  &3.8  &\nodata  &0.  \\
0.2442  &-19.01  &1.40  &0.  &21.89  &\nodata &0.8  &\nodata &\nodata  &5.1  &1.72  &2.  \\
0.2347  &-19.06  &1.47  &0.  &20.26  &\nodata &0.5  &\nodata &\nodata  &4.1  &\nodata  &3.  \\
0.2199  &-19.20  &1.03  &0.  &21.12  &\nodata &0.7  &\nodata &\nodata  &3.7  &\nodata  &1.  \\
0.2134  &-19.15  &1.50  &0.  &21.26  &\nodata &0.7  &\nodata &-5.2  &4.7  &1.53  &2.  \\
0.2099  &-19.26  &1.79  &0.28  &20.43  &17.77  &0.5  &1.1  &-2.4  &2.7  &1.72  &0.  \\
0.2068  &-19.32  &1.50  &0.07  &21.2  &10.77  &0.7  &-0.6  &\nodata  &4.4  &\nodata  &1.  \\
0.2393  &-19.14  &1.23  &1.  &\nodata &14.1  &\nodata &0.8  &\nodata  &6.5  &\nodata  &0.  \\
0.2658  &-19.15  &2.09  &1.  &\nodata &13.76  &\nodata &0.7  &\nodata  &0.8  &2.02  &0.  \\
0.2210  &-19.46  &2.04  &0.21  &21.55  &11.4  &0.7  &-0.2  &\nodata  &3.6  &\nodata  &1.  \\
0.2508  &-19.29  &1.21  &0.  &21.05  &\nodata &0.7  &\nodata &\nodata  &3.0  &2.07  &0.  \\
0.2279  &-19.33  &1.91  &1.  &\nodata &13.45  &\nodata &0.7  &9.4  &4.7  &2.12  &2.  \\
0.2478  &-19.28  &1.62  &0.45  &22.46  &13.93  &0.9  &0.5  &\nodata  &3.  &2.33  &0.  \\
0.2166  &-19.04  &1.20  &0.23  &21.9  &15.99  &0.8  &0.8  &-2.4  &8.3  &1.57  &3.  \\
0.2117  &-19.25  &1.98  &1.  &\nodata &13.31  &\nodata &0.7  &-2.6  &2.4  &2.39  &2.  \\
0.2427  &-19.07  &1.22  &0.  &19.81  &\nodata &0.5  &\nodata &-20.  &4.9  &1.67  &2.  \\
0.2224  &-19.63  &2.06  &0.35  &21.48  &8.95  &0.7  &-0.5  &\nodata  &0.7  &\nodata  &0.  \\
0.2223  &-19.46  &1.93  &1.  &\nodata &13.85  &\nodata &0.7  &-8.0  &5.7  &2.61  &2.  \\
0.1920  &-19.60  &1.23  &0.15  &21.69  &14.08  &0.8  &0.3  &\nodata  &3.0  &\nodata  &3.  \\
0.2374  &-19.48  &1.75  &1.  &\nodata &11.78  &\nodata &0.3  &-2.9  &2.8  &1.89  &2.  \\
0.2078  &-19.69  &1.55  &1.  &\nodata &12.12  &\nodata &0.4  &\nodata  &2.0  &\nodata  &4.  \\
0.2659  &-19.64  &2.01  &0.35  &20.71  &13.3  &0.6  &0.4  &\nodata  &-1.0  &2.08  &0.  \\
0.2202  &-19.64  &1.46  &0.11  &20.41  &9.97  &0.6  &-0.5  &\nodata  &3.7  &\nodata  &0.  \\
0.2593  &-19.55  &1.40  &0.1  &19.76  &10.26  &0.5  &-0.5  &\nodata  &1.6  &2.19  &1.  \\
0.2550  &-19.76  &0.95  &0.06  &21.45  &10.87  &0.8  &-0.5  &\nodata  &-0.6  &1.17  &3.  \\
0.1901  &-19.35  &1.47  &0.  &20.99  &\nodata &0.8  &\nodata &\nodata  &8.3  &1.48  &4.  \\
0.2047  &-19.50  &1.12  &0.  &20.04  &\nodata &0.6  &\nodata &-9.3  &12.6  &1.9  &0.  \\
0.2670  &-19.73  &1.29  &0.  &20.78  &\nodata &0.8  &\nodata &\nodata  &3.4  &1.51  &0.  \\
0.2461  &-19.94  &1.61  &0.27  &20.95  &13.72  &0.8  &0.5  &-56.  &-1.3  &2.02  &2.  \\
0.2185  &-20.11  &1.81  &0.  &21.29  &\nodata &0.9  &\nodata &-5.0  &4.7  &1.61  &1.  \\
0.2481  &-19.91  &1.77  &1.  &\nodata &15.72  &\nodata &1.5  &\nodata  &2.7  &3.17  &3.  \\
0.2676  &-20.26  &2.11  &0.44  &22.51  &13.18  &1.1  &0.5  &\nodata  &-0.4  &2.26  &2.  \\
0.2178  &-20.12  &1.56  &0.16  &20.53  &13.61  &0.8  &0.4  &0.9  &3.5  &2.03  &1.  \\
0.2271  &-20.51  &2.11  &1.  &\nodata &13.62  &\nodata &0.9  &\nodata  &0.9  &\nodata  &1.  \\
0.2551  &-20.30  &1.55  &0.76  &19.84  &14.77  &0.4  &1.  &-3.7  &6.9  &2.01  &3.  \\
0.2243  &-20.42  &1.87  &0.38  &21.99  &15.04  &1.0  &0.9  &\nodata  &1.2  &\nodata  &1.  \\
0.2658  &-20.42  &1.63  &0.11  &21.29  &9.49  &1.0  &-0.4  &-12.9  &5.3  &1.96  &0.  \\

\enddata
\end{deluxetable}

\clearpage
\normalsize
\centerline{References}
\vskip 5pt

Abraham R.G. et al, 1996, ApJ, 462, 96

Carlberg R.G. et al, 1996, ApJ, 462, 32

Hutchings J.B., and Neff S.G., 1991, AJ, 101, 434

Morris S.L. et al, 1998 ApJ, 507, 84

Saintonge A., Schade D., Yee H.K.C., Carlberg R., Ellingson E., 
2001 (in preparation)

Schade D., Lilly S.J., LeFevre O., Hammer F., Crampton D., 1996, ApJ, 464, 79

\clearpage
\centerline{Captions to Figures}

1. Spatial distribution of A2390 cluster galaxies in this paper. 
The Declination scale is
exaggerated for clarity.  Upper panel shows the MOS sample,
which was used in all other diagrams, separated into the 3 principal
morphology groups. The lower panel shows the other imaging samples.
These are less extensive but were used for independent checks on the
morphology results. 

2. Bulge to disk ratios for cluster and field galaxies for A2390.
Model errors in B/T are typically $\pm$0.08 or less.

3. Disk surface brightness and intrinsic-colour distributions for A2390 cluster
and field galaxies. The dotted lines are the mean values for the field sample.
Note the comparative lack of low surface brightness disks in the blue 
cluster galaxies. Model error bars typically 0.2 magnitudes, or about twice
the size of the dots.

4. Disk scale-length changes with galaxy luminosity for A2390 and field
galaxies. The dotted line is the linear fit to the field distribution, and
the dashed line is the parabola fit to the cluster sample. Scale length
error bars are typically 15\%, about 0.1 in the log scale plotted. 

5. The fraction of cluster galaxies by morphology type for 4 bins of
local galaxy density. The galaxy density is strongly correlated with
radius within the cluster, as indicated by the R$_{norm}$ values.
The middle two bins are where there is significant degeneracy between
radius and local density, as result of galaxy substructure, and have equal
numbers of galaxies.

6. Spectroscopic quantities (from Abraham et al 1996) as a function of
bulge to disk ratio for A2390 cluster and field galaxies. 

7. Spectroscopic quantites for A2390 cluster and field galaxies as function
of 5-point interaction scale. The circles show the mean values for each group,
and the percentages in the right panel refer to the full numbers of the
cluster and field samples.

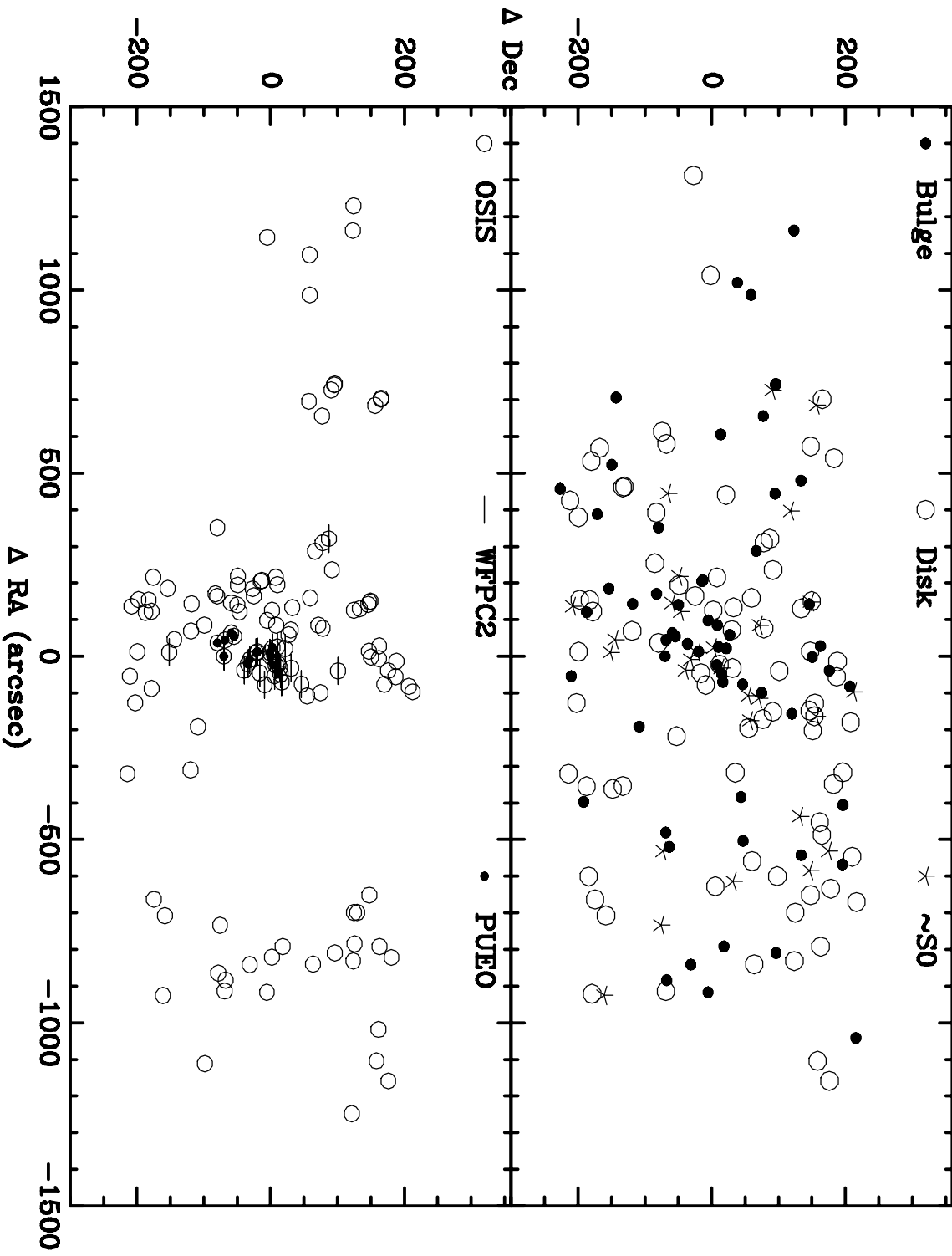
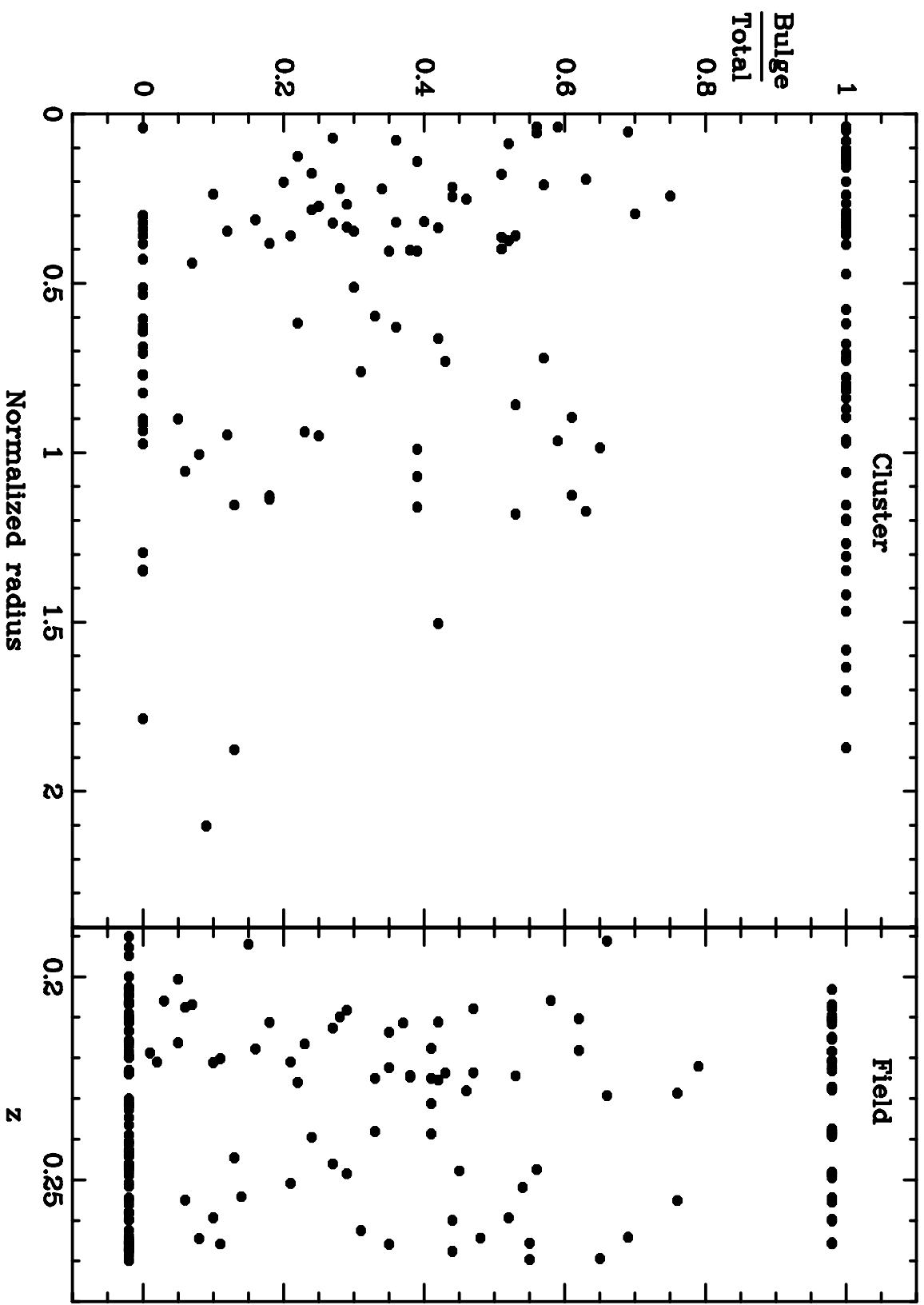
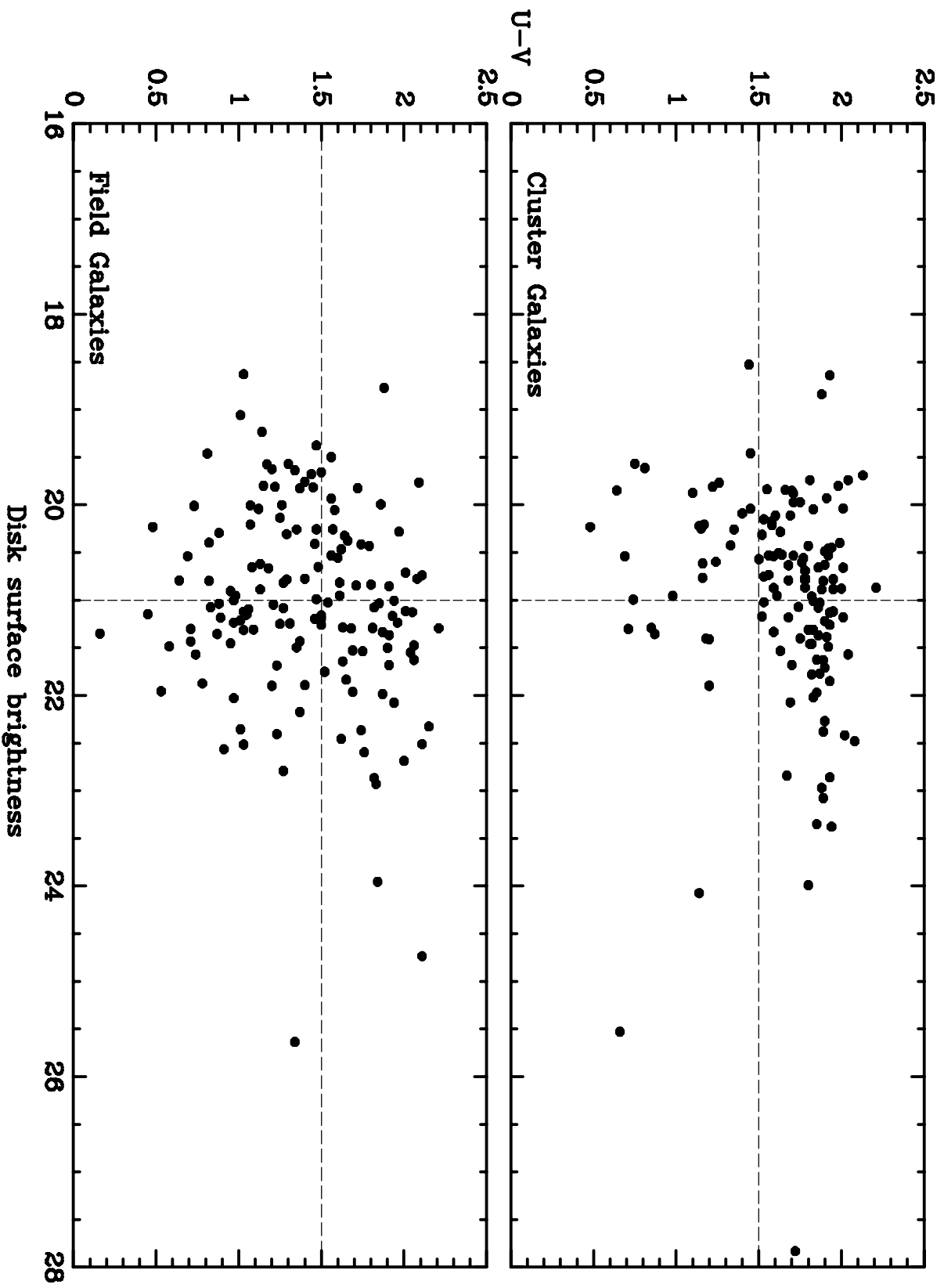
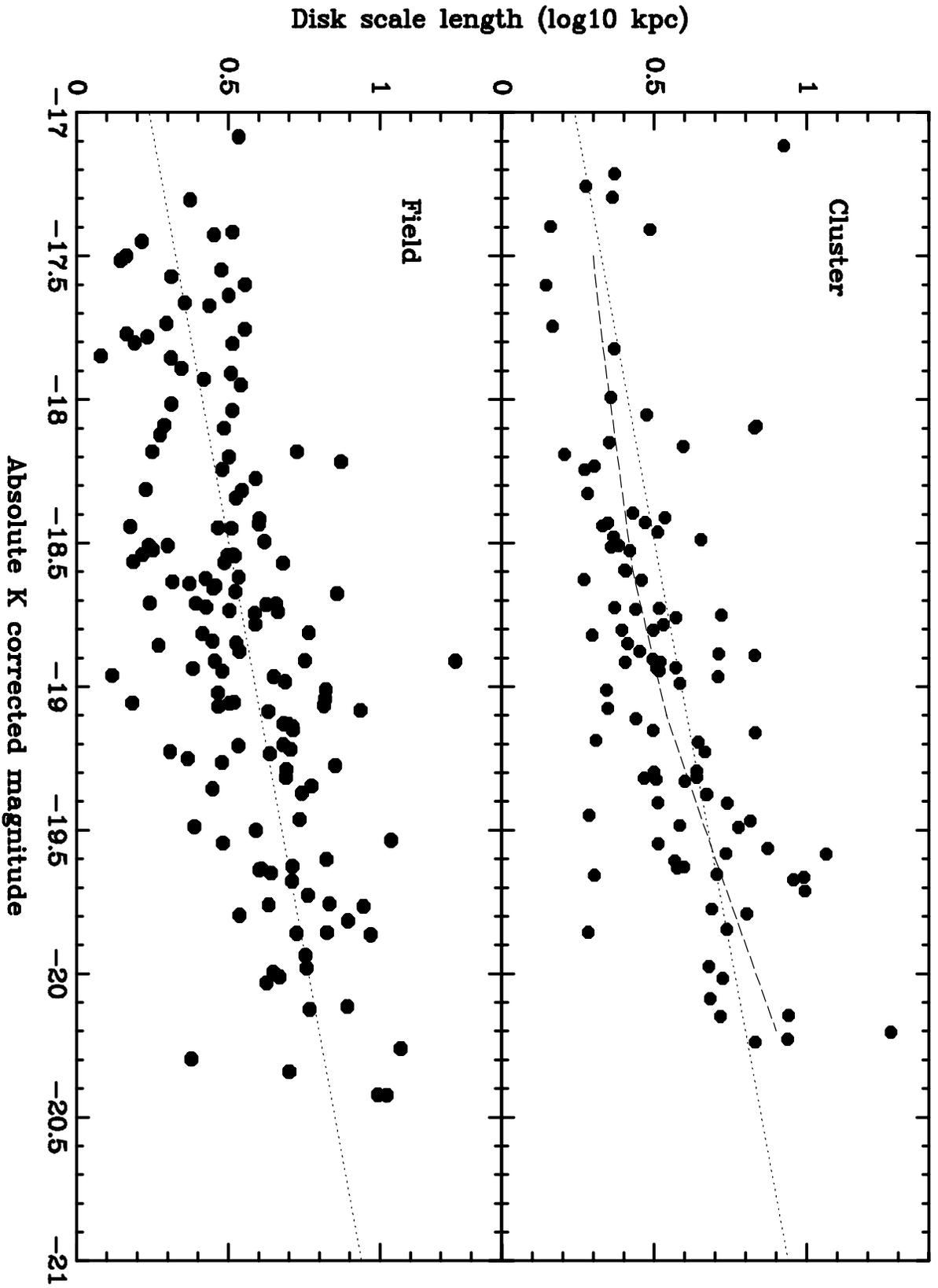
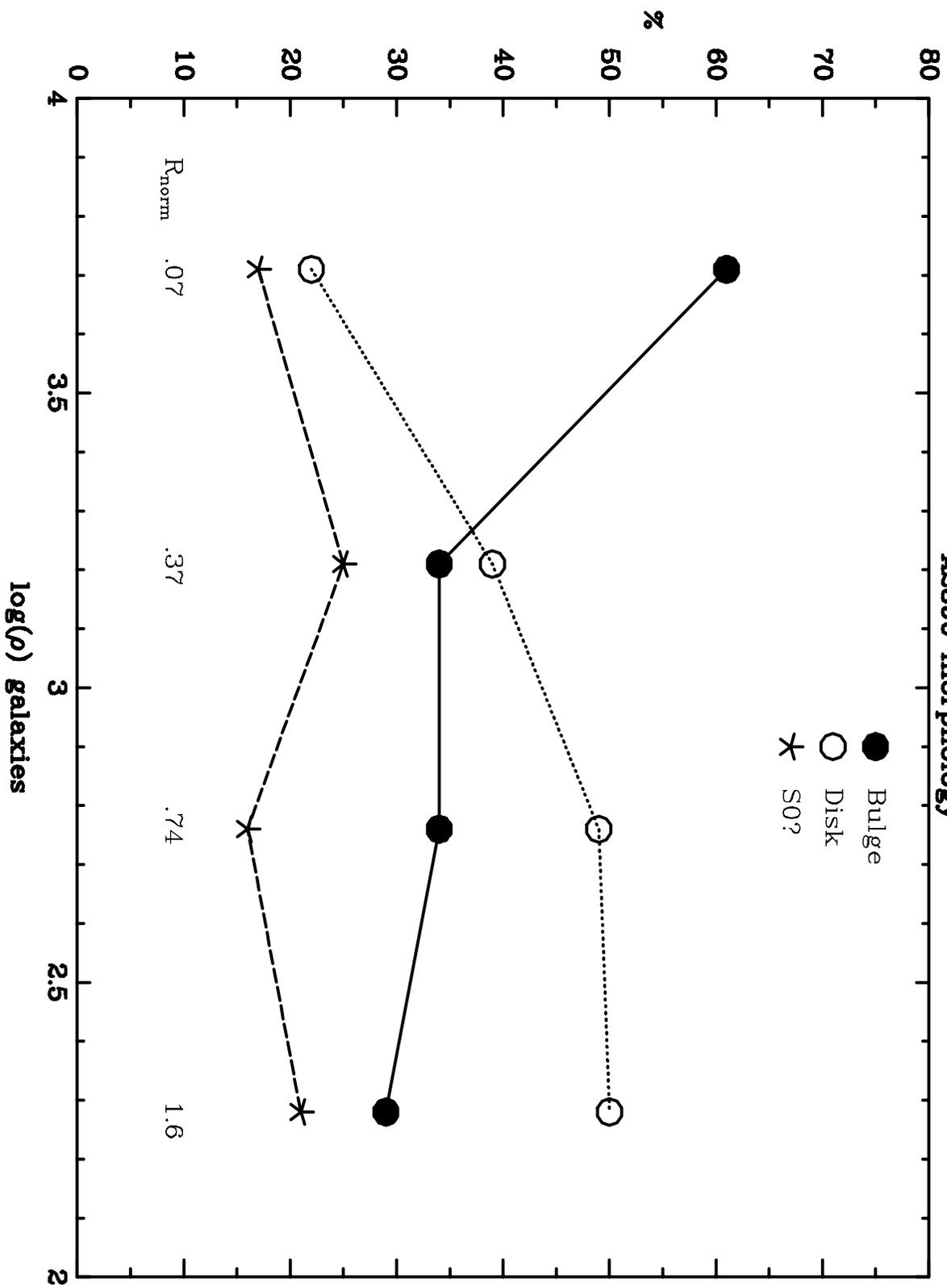
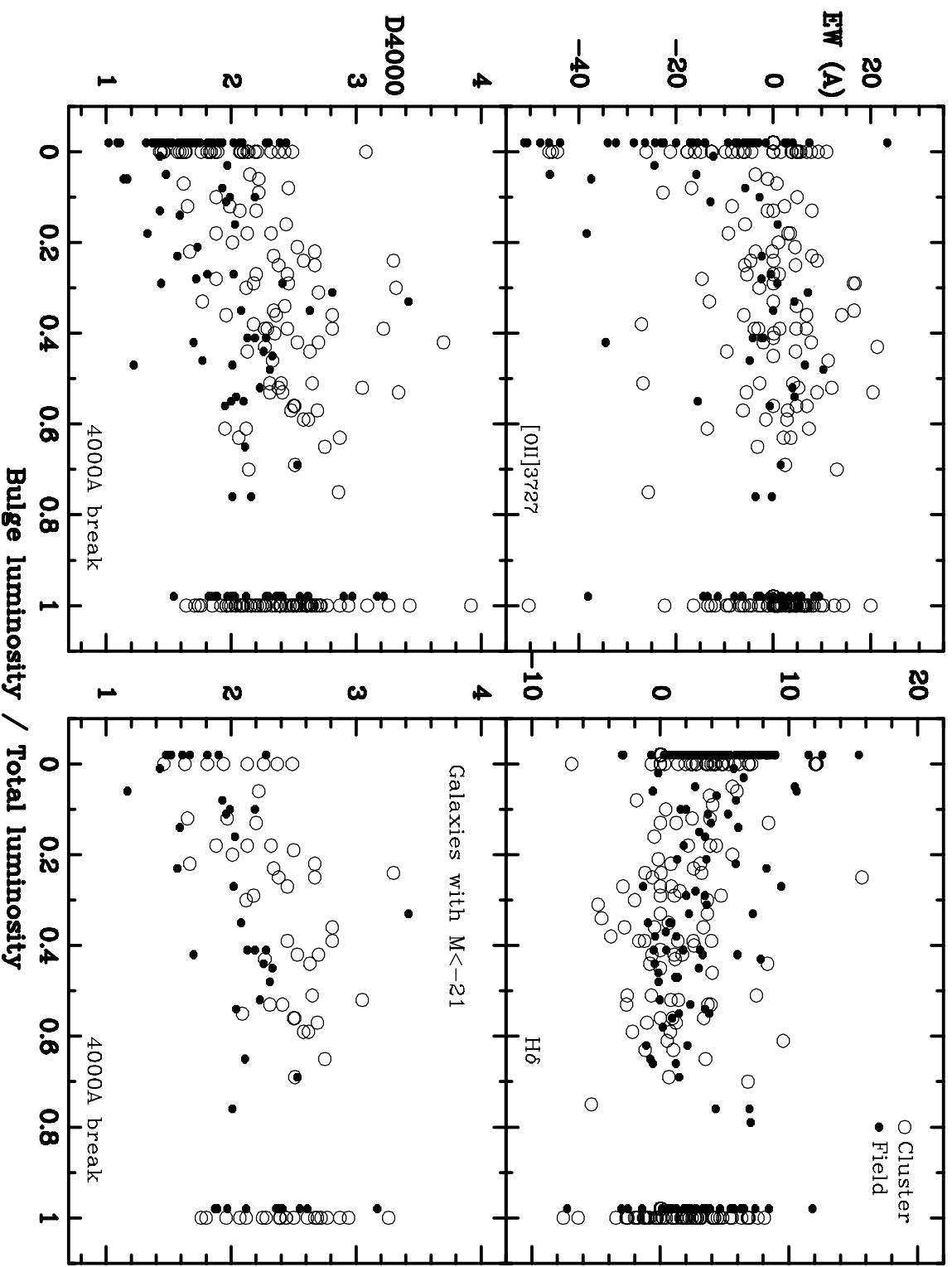
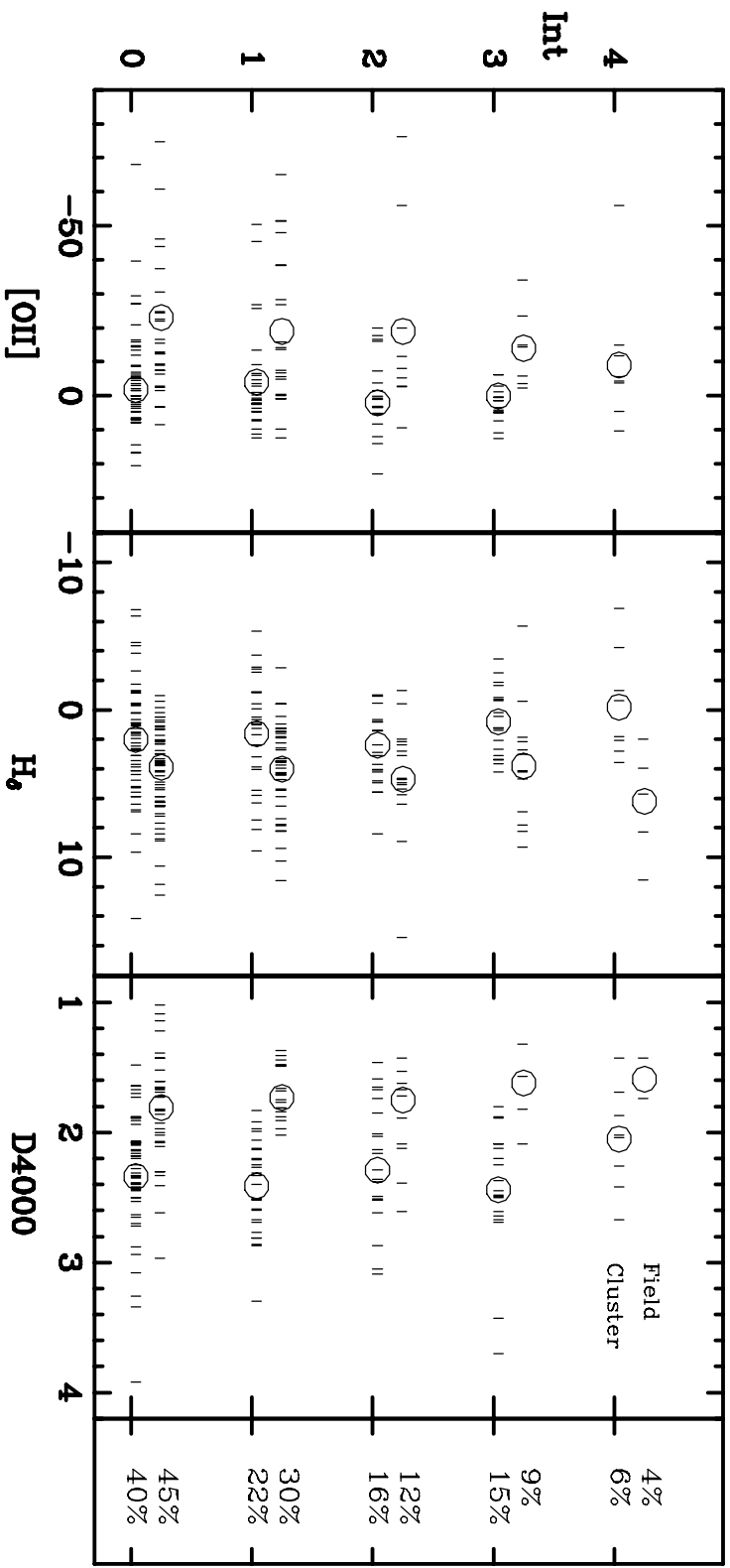
\end{document}